\documentclass[twocolumn,aps,pra]{revtex4}

\usepackage{amsmath,amsfonts,amssymb}
\usepackage{wrapfig}
\usepackage{graphicx}
\usepackage{bbm}
\usepackage{color}
\usepackage{float}
\usepackage{subfigure}
\usepackage{textcomp}
\usepackage[english]{babel}
\usepackage{pifont}

\usepackage{mathrsfs}
\usepackage{wasysym} 
\usepackage{psfrag}
\usepackage[colorlinks=true,linkcolor=blue,citecolor=blue]{hyperref}
\usepackage{enumerate} 

\newcommand{\be}{\begin{equation}}
\newcommand{\ee}{\end{equation}}
\newcommand{\bea}{\begin{eqnarray}}
\newcommand{\eea}{\end{eqnarray}}
\newcommand{\ba}{\begin{array}}
\newcommand{\ea}{\end{array}}
\newcommand{\bc}{\begin{center}}
\newcommand{\ec}{\end{center}}
\newcommand{\ben}{\begin{enumerate}}
\newcommand{\een}{\end{enumerate}}
\newcommand{\bi}{\begin{itemize}}
\newcommand{\ei}{\end{itemize}}
\newcommand{\bt}{\begin{table}}
\newcommand{\et}{\end{table}}
\newcommand{\btab}{\begin{tabular}}
\newcommand{\etab}{\end{tabular}}
\newcommand{\bfi}{\begin{figure}}
\newcommand{\efi}{\end{figure}}

\newcommand{\bd}{\begin{description}}
\newcommand{\ed}{\end{description}}

\newcommand{\nn}{\nonumber}

\newcommand{\tc}{\textcircled}

\newcommand{\abs}[1]{\left|#1\right|}

\makeatletter

\def\compoundrel#1\over#2{\mathpalette\compoundreL{{#1}\over{#2}}}
\def\compoundreL#1#2{\compoundREL#1#2}
\def\compoundREL#1#2\over#3{\mathrel
      {\vcenter{\hbox{$\m@th\buildrel{#1#2}\over{#1#3}$}}}}
\makeatother

\usepackage{babel}
\usepackage{wasysym}
\usepackage{harmony}
\usepackage{semtrans}

\makeatother

\def\be{\begin{eqnarray}}
\def\ee{\end{eqnarray}}
\def\bee{\begin{eqnarray*}}
\def\eee{\end{eqnarray*}}

\begin{document}

\title{Projective simulation for classical learning agents: a comprehensive investigation}

\author{Julian Mautner$^{1,2}$, Adi Makmal$^{1,2}$, Daniel Manzano$^{1,2,3}$, Markus Tiersch$^{1,2}$ and Hans J. Briegel$^{1,2}$}

\affiliation{$^1$Institut f{\"u}r Theoretische Physik,
Universit{\"a}t Innsbruck, Technikerstra{\ss }e 25, A-6020 Innsbruck\\
$^2$Institut f{\"u}r Quantenoptik und Quanteninformation der
\"Osterreichischen Akademie der Wissenschaften, Innsbruck, Austria\\
$^3$Instituto Carlos I de Fisica Te\'orica y Computational,
University of Granada, Granada, Spain
}

\date{\today}

\begin{abstract}
We study the model of \emph{projective simulation} (PS), a novel approach to artificial intelligence based on stochastic processing
of episodic memory which was recently introduced \cite{BriegelCuevas12}.
Here we provide a detailed analysis of the model and examine its performance, including its achievable efficiency, its learning times and the way both properties scale with the problems' dimension. In addition, we situate the PS agent in different learning scenarios, and study its learning abilities. A variety of new scenarios are being considered, thereby demonstrating the model's flexibility.
Further more, to put the PS scheme in context, we compare its performance with those of Q-learning and learning classifier systems, two popular models in the field of reinforcement learning. It is shown that PS is a competitive artificial intelligence model of unique properties and strengths.
\end{abstract}

\maketitle

\section{Introduction}
Artificial intelligent agents are playing an increasingly important role in our modern life. Different from ordinary computers, 
intelligent agents are designed to operate autonomously in complex and possibly unknown environments. Intelligence is thereby 
understood as the agent's capability of acting in its environment in a rational and flexible way that maximizes its chance of 
success \cite{RusselNorvig03}. Applications of artificial agents include robots that interact with humans, operate in remote space, 
or search the internet (netbots), while in biology, the study of artificial agents may also provide new perspectives to model animal 
behavior. Comprehensive introductions can be found in modern textbooks  \cite{RusselNorvig03,PfeifferScheier99,FloreanoMattiussi08}.

In this paper we study a novel approach to \emph{artificial intelligence} (AI) which was first introduced recently \cite{BriegelCuevas12} and which we call \emph{projective simulation} (PS).  
Projective simulation constitutes a model of information processing for artificial agents, in which the agent effectively projects itself into potential future scenarios, according to its previous experience. 
The model is based on simple stochastic processes, thus providing a physically grounded approach toward an embodied agent design. The model can be naturally applied to problems in reinforcement learning, where the agent learns 
via interaction with some rewarding environment \cite{SuttonPhD84,SuttonBarto98}. At the same time, the notion 
of PS is more general and can also be seen as a principle and building block for complete agent architectures and 
computational intelligence  \cite{Poole}.

A central component of the PS scheme is a specific type of memory, which we denote as \emph{episodic \& compositional 
memory} (ECM).
The ECM is structured as a directed, weighted network (graph), and we refer to each node of this network as a \emph{clip}.
The ``clips" are the basic units of memory and correspond to short episodic experiences. They consist e.g.\ of remembered 
percepts,
or actions, or simple combinations thereof. In everyday life, examples of such clips could be ``seeing a red ball'', ``kicking a ball'', 
or its combination ``seeing a red ball and kicking it''. A clip in ECM can be excited through some perceptual stimulus from the 
environment, and this excitation then hops to an adjacent clip with probability that is correlated to the strength of the directed 
edge between these two clips. Perceptual input thereby leads to a \emph{random walk through memory}, organized as a network 
of clips. 
This random walk finally reaches its end once the excitation of a so-called ``action clip" couples out -- also probabilistically in general -- and leads to a corresponding real action of the agent in its environment.

``Learning" is effectively achieved through dynamic changes of the ECM network.
The network is continuously adjusted, through experience, according to rewards obtained from the environment.
This adjustment may take place in terms of both the clips themselves as well as the weights of the edges connecting different clips.
At the beginning, the PS agent is situated in an environment as a \emph{tabula rasa}, meaning that its ECM network shows no preferences toward any kind of action, or behavior\footnote{Clearly, the mere structure of the ECM network already encodes hard-wired information, e.g., regarding the potential percepts and actions the agent may encounter and perform. From this perspective, no agent is a perfect \emph{tabula rasa}.}. Then, in subsequent time steps, the agent's actions are rewarded by the environment. These rewards are directly translated into corresponding changes within the ECM according to simple prescribed rules.
The changes in the ECM may then result in the PS agent taking different actions, thus initiating a new feedback loop.
A snapshot of the ECM at each time step thus reflects the past experience of the PS agent with respect to its actions and rewards. When successful, the resulting ECM allows the PS agent to take actions that maximize, on average, the rewards it obtains from the environment.

We note that the model of projective simulation can also 
be generalized to quantum mechanical operation. This leads to the concept of a \emph{quantum agent} which employs the principles of quantum mechanics, such as quantum superposition and parallelism, for processing its episodic 
memory \cite{BriegelCuevas12}. Even though it is not the topic of this paper, the possibility of a quantum generalization can be seen as a unique feature of the model. Research along this line will be published in a separate work.


This paper provides, in addition to a thorough analysis of the PS model, three main contributions:
First, the PS agent is situated in a variety of novel scenarios, each of which confronts the agent with a different kind of challenge, thereby enabling us to evaluate the agents learning abilities and demonstrate its flexibility. Second, a new ``glowing" mechanism is introduced to the model, which builds up correlations between ECM excitations that were activated at different times. This is shown to have a dramatic effect on the PS performance in certain scenarios. Last, we conduct a detailed comparison between the PS model and two well established approaches to reinforcement learning problems, namely, Q-learning and learning classifier systems (see below). This comparison is twofold: on one, more technical, level, the models are repeatedly compared with respect to their qualitative performance, 
whereas on a more conceptual level, the models are compared with respect to their simplicity, an important trait of any artificial intelligent agent and of embodied agents in particular.

The paper is structured as follows: section \ref{sec:Basic_notions_and_features}
is devoted to formal aspects of PS, basic notions and features. In particular, 
we start in section \ref{sec:PS_formalism} with a formal description of PS; 
Afterwards, in section \ref{toyModel} we introduce our simplest toy model, for which we analyze the performance of PS \emph{analytically} in section \ref{sec:approxSol}, where closed expressions are obtained for the asymptotic efficiency of the PS agent and the initial slope of its learning curve; In section \ref{sec:damping} we then describe the role of damping in PS; Last, we define a notion of ``learning time" in section \ref{sec:scaling}, and study its scaling properties when increasing the number of inputs and/or actions.

In sections \ref{sec:tempCorr}-\ref{sec:composition}, we then examine the ability of the PS agent to handle more advanced kinds of scenarios, where each scenario is of completely different nature. We show, for each case, how simple changes in the update rules of the ECM may lead to better performances, thereby making the agent more flexible. 
In particular, section \ref{sec:tempCorr} considers scenarios in which rewards may depend not only on present actions, but also on actions taken in the \emph{past}. Such scenarios, to which we refer as \emph{temporal correlations}, require some mechanism to correlate between actions taken at different times. We show that the PS scheme can be extended to account for such correlations, by allowing a slower decay of the excitations in the ECM, which we denote as \emph{afterglow}.
Next, in section \ref{asso} we study scenarios in which it is beneficial to
notice similarities between inputs and to associate the corresponding percepts. We show how the ECM can be automatically shaped to allow for such associations between percept-clips, and denote this capability as \emph{associative memory}.
Last, in section \ref{sec:composition} we aim at scenarios where composite actions are needed, i.e.\ where it is beneficial to combine known actions into new ones. We show how the PS scheme allows the agent to ``try out" such composite actions within the ECM. These new action clips are then available to the agent, increasing its adapting and learning capabilities.

To put the PS scheme in perspective with respect to existing models of AI, we
perform a detailed comparison throughout the paper between PS and other AI models, whenever such a comparison is sensible.
Out of many possible existing AI schemes we chose to focus on two representatives, namely Q-learning (QL) 
\cite{WAT,Watkins89,SuttonBarto98,RusselNorvig03}  
(with and without Dyna-style planning \cite{Sutton90,SuttonEtAl08} extension) which is a well known model in reinforcement learning, and on extended learning classifier systems  (XCS) \cite{Wilson95} that is an advanced variant of standard learning classifier systems (LCS)  \cite{Holland75,BullKovacs05,UrbanowiczMoore09}. 
Both are shortly described in appendices \ref{appx_QL} and \ref{appx_LCS}, respectively.
We chose to focus on these two models for two main reasons: first, both models are popular and well studied, and second, because they are known to perform well on our set of learning scenarios. In particular, QL is used as a reference model in the context of ``temporal correlations" (section \ref{sec:tempCorr}), whereas the XCS is used for comparison in the context of  ``associative memory" and ``composite actions" (sections \ref{asso} and \ref{sec:composition}).

Last, in section \ref{sec:comparison} we study the PS model in terms of its \emph{simplicity}, that is, we look
at the resources it requires, and estimate their complexity. This is an important aspect, as PS ultimately aims at providing a framework of embodied agent design \cite{PfeifferScheier99}, grounded only on physical processes, rather than computational ones. Here we study the essential resources of PS in terms of required parameters, basic data structure, and inherent processes. The PS is then compared to both QL and XCS in this context.
Section \ref{sec:conclusion} concludes the paper.

\section{Basic notions and features} \label{sec:Basic_notions_and_features}
The PS agent is conceived as an entity situated in a (partially unknown) environment, which
receives inputs via its sensors and can perform different actions. The actions of the agent are rewarded by the environment,
which affects the internal structure of its memory. The PS agent has, however, no explicit model of the environment which predicts the next state or reward and in that sense the PS is ``model-free" \cite{SuttonBarto98}.
From the point of view of an external observer, the agent may be described by a conditional probability $P^{(t)}(a|s)$ of performing an action $a$ given a percept $s$ (denoted elsewhere as the agent's ``stochastic policy" \cite{RusselNorvig03}). Yet, a complete description of the agent connects $P^{(t)}(a|s)$ with the internal 
state of the agent's memory at time $t$, and specifies how its memory is modified as the agent interacts with the environment.

The model of projective simulation provides such a description in terms of stochastic processes, as will be specified in the next subsection.

\subsection{The PS formalism \label{sec:PS_formalism}}
In what follows we list the main formal points that constitute the PS model:
\begin{itemize}
    \item \textbf{The ECM} is the central component of PS, defined as a directed, weighted network (graph). Each node of this network is called \emph{a clip}.
	 \item \textbf{Clips} represent fragments of episodic experiences, which are defined as $L$-tuples $c = (c^{(1)},c^{(2)},...,c^{(L)})$.
Each of $c^{(i)}$ is an internal representation of either a percept (${\cal S}$), or an action (${\cal A}$), where both are defined below.
In this paper, we consider only clips composed of length $L=1$.
\item \textbf{Percepts (``inputs")} are defined as $N$-tuples $s=(s_1,s_2,...,s_N) \in {\cal S} \equiv {\cal S}_1 \times {\cal S}_2 \times ... \times {\cal S}_N,s_i \in \{1,...,|{\cal S}_i|\}$, where the number of possible percepts is given by $S\equiv|{\cal S}| = |{\cal S}_1| \cdots |{\cal S}_N|$. The structuring of the input into subspaces is usually naturally given, for example when considering robots: ${\cal S}_1$ might represent a visual sensor,  ${\cal S}_2$ may account for an acceleration sensor, and so forth. The notation in the original proposal~\cite{BriegelCuevas12} differentiates between the immediate physical percept (sensory input) caused by the environment denoted by $s$ and its representation in memory, i.e.\, a percept clip denoted by \tc{$s$}. Here we employ the notation $s$ both for the percept and for the percept clip, as long as there is no danger of confusion.
	\item \textbf{Actions (``outputs")} are given as $M$-tuples: $a=(a_1,a_2,...,a_M) \in {\cal A} \equiv {\cal A}_1 \times {\cal A}_2 \times ... \times {\cal A}_M,a_i \in \{1,...,|{\cal A}_i|\}$, where the number of possible actions is given by $A\equiv|{\cal A}| = |{\cal A}_1| \cdots |{\cal A}_M|$.
The action space is also structured using subspaces, which could be, e.g.\, moving, beeping, touching, etc. Similar to percepts, real actions $a$ are conceptually different from remembered actions, i.e.\, action clips, which were denoted by \tc{$a$} in the original proposal \cite{BriegelCuevas12}. Again, in the present work we will employ the notation $a$ both for the real action and for the action clip.
	\item \textbf{Each edge}, connecting clip $c_i$ to clip $c_j$, has a dynamic weight $h^{(t)}(c_i,c_j)$, which changes over time, and is denoted as the \emph{$h$-value} of this edge.
Initially, at time $t=0$, there exist edges directed from each percept-clip to each action-clip. The $h$-values of the edges are initialized to $h^{(0)}(c_i,c_j)=1$, and obey $h^{(t)}(c_i,c_j) \geq 1$ at each time $t$.
	\item \textbf{The hopping probability} according to which an excitation hops from clip $c_i$ to clip $c_j$ is given by $p^{(t)}(c_j|c_i) = \frac{h^{(t)}(c_i, c_j)}{\sum_k h^{(t)}(c_i, c_k)}$ where the sum is over all clips $c_k$ connected to clip $c_i$.
\item \textbf{``Emotion" tags} are degrees of freedom which can be assigned internally to percept-action edges to indicate e.g.\ whether or not the corresponding transition has been rewarded the last time it was taken. 
These tags can be used to memorize the most recent reward on a given transition, thereby enabling the detection of short-time changes in the environment. 
Formally, the emotion tags are given by $e(s,a)=(e_1,e_2,...,e_k) \in {\cal E} \equiv {\cal E}_1 \times {\cal E}_2 \times ... \times {\cal E}_k, e_i \in \{1,...,|{\cal E}_i|\}$, where here we restrict ourselves to the case where $k=1$ and $e(s,a) \in {\cal E} = \{ \smiley, \frownie \}$.
 	\item \textbf{Reflection} is the mechanism that exploits the emotion tags: before an action $a$ is coupled out as a response to an excitation of percept clip $s$, the emotional tag $e(s,a)$ is checked. If it is
positive ($\smiley$) the action is performed, but if it is
negative ($\frownie$) the random walk process is restarted. This allows the agent to reconsider, so to speak, its choice. The maximum number of random walks per decision is limited by a predefined ``reflection time" parameter $R$, whose default value $R=1$ means \emph{no reflection}.
	\item \textbf{The interface} between the PS agent and the external environment is realized via its sensors and actuators and their connection to memory. An external percept $s$ excites a certain percept clip $c$ according to ${\cal I}(c|s)$, an input-coupling probability function. Similarly, an action-clip couples out to perform a real action $a$ according to an output-coupling probability function ${\cal O}(a|c)$. The coupling functions connect the internal random walk, described by the hopping probabilities $p^{(t)}(c_j|c_i)$, with the external behavior of the agent, described by $P^{(t)}(a|s)$. 
To be more explicit, in the simplest case, the probability that a percept $s$ will initiate a specific random walk through clips \{$c_{k_1}\rightarrow c_{k_2}\rightarrow\ldots\rightarrow c_{k_l}\}$ (including possible repetitions), 
which will eventually lead to action $a$, is given by ${\cal O}(a|c_{k_l})\Pi_{j=1}^{l-1}p^{(t)}(c_{k_{j+1}}|c_{k_j})  {\cal I}(c_{k_1}|s)$.
The external conditioned probability $P^{(t)}(a|s)$ is then given by summing up the probabilities over all such possible paths inside the clip network. 
Here we consider only the case where ${\cal I}(c|s)$ and ${\cal O}(a|c)$ are simple Kronecker delta functions\footnote{This is consistent with our restriction to clips of length $L=1$, which means that $c$ is itself a single remembered percept or action.}. 
This corresponds to the simplification that a percept always excites the corresponding percept-clip and an action-clip always couples out to produce the corresponding action.
\end{itemize}

The basic process underlying the PS model is a stochastic one.
Each time step $t$ begins with a percept coming from the environment (an input) and exciting a memory clip $c_i$ inside the network according to ${\cal I}(c|s)$. Next, the excitation hops from clip $c_i$ to one of its neighboring clips, $c_j$, with the hopping probability $p^{(t)}(c_j|c_i)$. This hopping process then continues in a random walk manner, allowing the excitation to propagate through the clip network. The hopping process then reaches its end, once an action clip is encountered and couples out to an action in the real-world according to ${\cal O}(a|c)$.

When an action is performed, the agent gets a reward $\lambda \geq 0$ from the environment. As a result, the $h$-values of all edges are updated according to the following two rules: (i) The $h$-values of all \emph{activated} edges, i.e.\ edges that were traversed during the last random walk, are updated according to
\begin{equation} \label{eq:updateRuleOrgI}
h^{(t+1)}(c_i, c_j) = h^{(t)}(c_i, c_j) - \gamma (h^{(t)}(c_i, c_j) - 1) +  \lambda,
\end{equation}
where $t$ is the current time step, $\gamma$ ($0 \le \gamma \le 1$) is a damping parameter (see also section \ref{sec:damping}), and $\lambda$ quantifies the reward given by the environment.
(ii) The $h$-values of all \emph{other} edges of the network are merely damped, described by the rule
\begin{equation} \label{eq:updateRuleOrgII}
h^{(t+1)}(c_i, c_j) = h^{(t)}(c_i, c_j) - \gamma (h^{(t)}(c_i, c_j) - 1) .
\end{equation}
This update of the ECM network concludes a single time step.

At each time step all the $h$-values are therefore reduced, by a factor of $\gamma (h-1)$, whereas only edges that were actively used in the last step, are given the reward $\lambda$. When the PS-agent obtains a positive reward, the edges that were visited during the random walk that led to the correct action are then strengthened, thereby increasing the probability that they will be used again in the future. On the other hand, when a wrong action is taken and no reward is given, i.e.\ $\lambda = 0$, all edges are merely damped, including those that were used in the current step, thus reducing the probability to use them in the future.

Last, we remark that Eq.\ (\ref{eq:updateRuleOrgI})-(\ref{eq:updateRuleOrgII}) is a simple, yet obviously not the only possible update rule for the network, where different behaviors may emerge from choosing different update rules. This makes PS a very flexible framework. Throughout this paper, this flexibility will be examined and demonstrated.

\subsection{Toy model: The invasion game\label{toyModel}}
For an initial analysis of the PS model we use a game we call the ``invasion game" \cite{BriegelCuevas12} as our simplest toy model (for more advanced scenarios see sections \ref{sec:tempCorr}-\ref{sec:composition}).
The game is composed of an attacker, a defender and several doors.
The agent has the role of the defender, whose task is to block the attacker.
At the beginning of each time step, the attacker and defender are facing each other at the same door. The attacker then shows a symbol indicating where it will appear next (e.g.\ left or right arrow), and the defender makes a move.
The defender has thus to learn the meaning of the symbols to take the correct action. If it managed to move to the correct door and block the attacker, it will receive a reward $\lambda>0$, whereas no reward is otherwise obtained.
The percept space here is a set of symbols indicating the direction of the attacker's motion. In the simplest version, where the attacker has only two symbols to show, this would amount to e.g.\ ${\cal S} = \{\leftarrow, \rightarrow \}$.
The action space is a set of possible actions. In the simplest case, only two actions are allowed: one step to the left and one to the right, therefore ${\cal A} = \{-,+\}$.

\begin{figure}[htb]
	\begin{center}
		\begin{minipage}{9cm}
				\includegraphics[width=8cm]{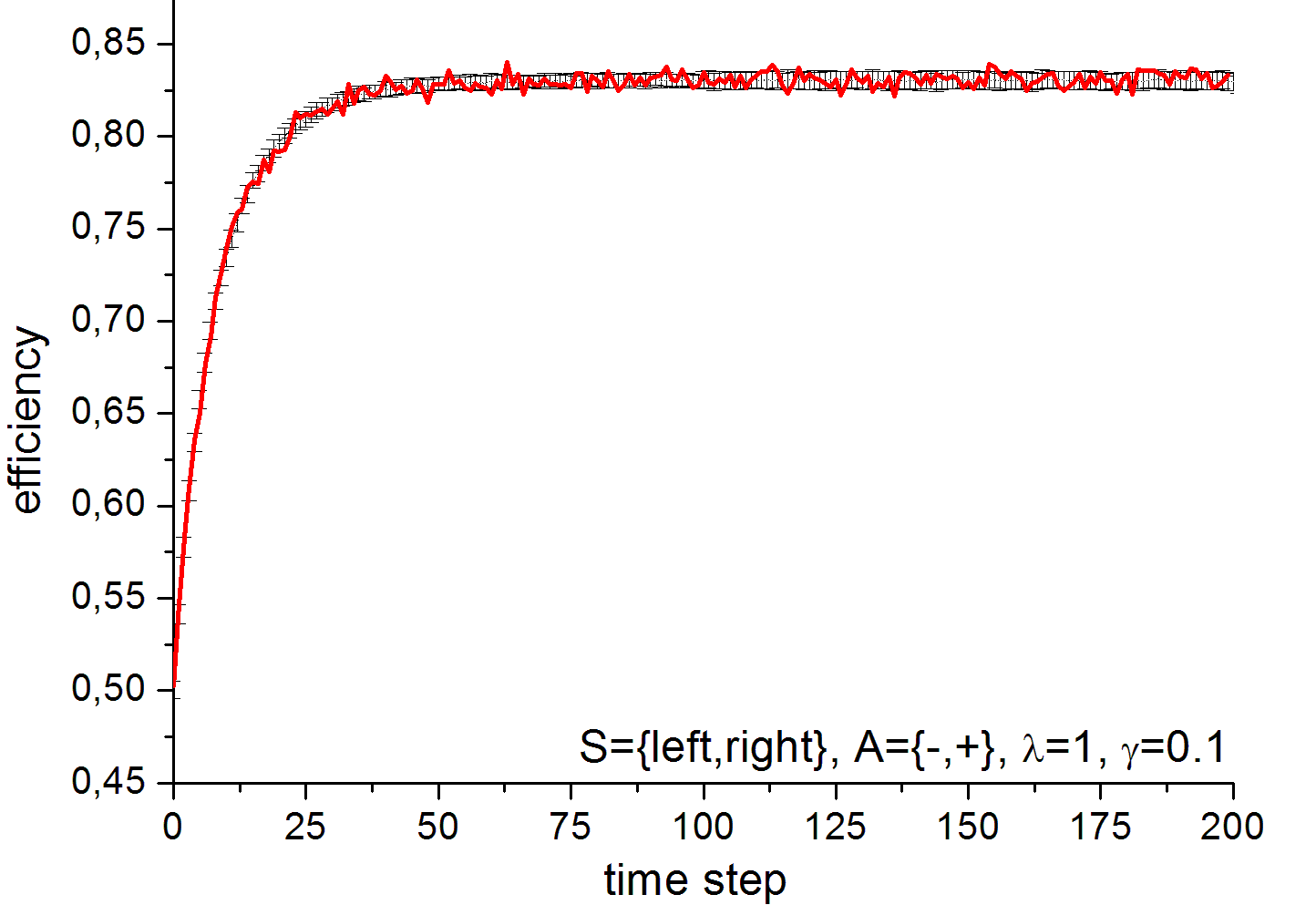}
			\end{minipage}
		\end{center}
	\caption{A typical learning curve of PS for the invasion game shows the efficiency, averaged over $10^4$ single agents, as a function of time steps (in red). Error bars on top of the curve were calculated using 100 different learning curves (using a total of $10^6$ runs) and they are of $2\sigma$ width (see text). It is seen that the errors are of the same order as the fluctuations.}
	\label{fig:errorBarsPlot}
\end{figure}

\begin{figure}[htb]
	\begin{center}
		\begin{minipage}{9cm}
				\includegraphics[width=8cm]{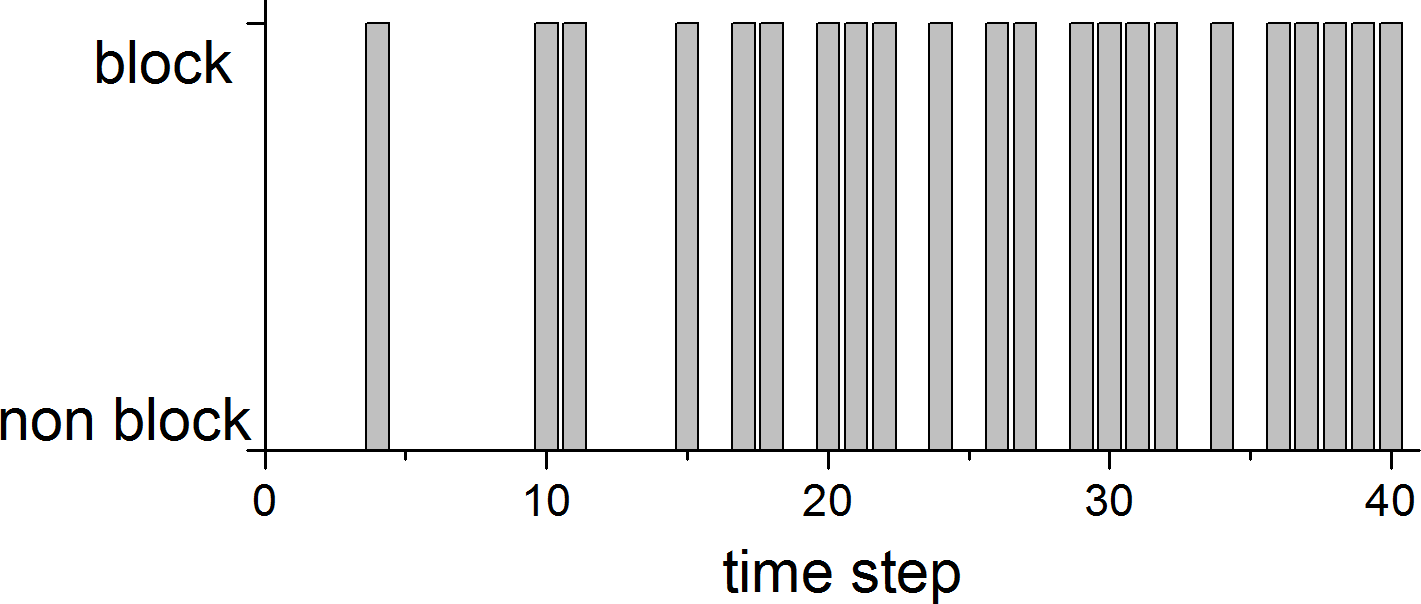}
			\end{minipage}
		\end{center}
	\caption{A trajectory of a single PS agent for 40 time steps of the invasion game. It is seen that the percentage of blocking events increase with time, indicating that the agent learns.}
	\label{fig:singleTrajectory}
\end{figure}

A typical learning curve of the PS in the invasion game is depicted in red in Fig.~\ref{fig:errorBarsPlot}.
The game is played by an ensemble of $Y$ agents, for $T$ time steps. At time $t=0$ all agents are similarly initialized to show no preference toward any action (their $h$-values are all set to 1). Then, at every time step, each of the agents is confronted with a random symbol, takes an action, and is possibly rewarded, followed by the update of its ECM according to Eq.\ \eqref{eq:updateRuleOrgI}-\eqref{eq:updateRuleOrgII}. A particular network will then develop for each of the individual agents, resulting with a different blocking efficiency of the different agents.
A trajectory of a single agent consists of a sequence of blocking and non-blocking actions, as shown in Fig.~\ref{fig:singleTrajectory}.
In what follows, we always use at least $Y=10^4$ agents, unless stated otherwise. We further define the  efficiency of each time step as the ratio $\frac{X}{Y}$, where $X$ is the number of agents that blocked the attacker at this time step. An efficiency of $0.85$ therefore means, that out of the $10^4$ agents, 8500 blocked the attacker, and 1500 did not.

To make sure that the choice of $Y=10^4$ agents is sufficient for meaningful statistics, we also calculated the error bars, as shown in Fig.~\ref{fig:errorBarsPlot}.
To that end, we calculated 100 efficiency curves (with each being averaged over $10^4$ agents).
Then out of this ensemble we calculated the averaged efficiency and the standard deviation for each time step. The error bars we show are of $2\sigma$ width, centered at the mean efficiency value. Fig.~\ref{fig:errorBarsPlot} shows that the errors are of the same order as the fluctuations. Therefore, in following plots, we omit the error bars, as they are indirectly given by the fluctuations.

\subsection{The PS learning curve: A heuristic analysis \label{sec:approxSol}}
In this section, we analyze the learning curve of PS
(such as shown, e.g.\, in Fig.~\ref{fig:errorBarsPlot}) from an analytical perspective. 
In particular, we develop closed and compact expressions which approximate the initial slope of the efficiency and its asymptotic value. These are most relevant properties: the initial slope provides a first indication for the agent learning time, whereas the asymptotic efficiency indicates the averaged maximal efficiency that can be achieved.

\begin{figure}[htb]
	\begin{center}
		\begin{minipage}{9cm}
				\includegraphics[width=8cm]{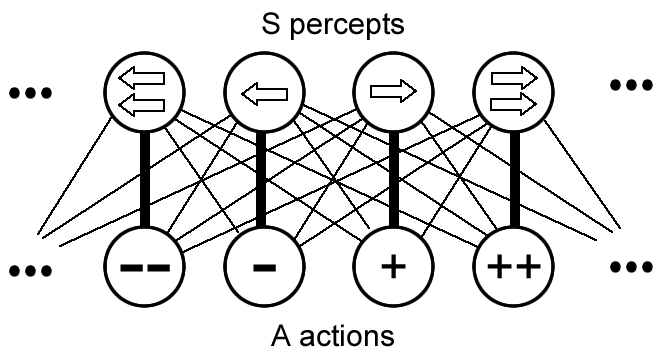}
			\end{minipage}
		\end{center}
	\caption{
A schematic illustration of the PS clip network, as built for the restricted invasion game that is used for our analytical derivation. Percept and action clips are shown on top and on the bottom, respectively, and thick (thin) edges denote rewarded (unrewarded) edges. Here, each percept clip has a single, unique, rewarded action. The resulting clip network is highly structured (though at each time step the strengths of the rewarded edges slightly vary).}
	\label{fig:clipNetworkForAnalytics}
\end{figure}

Obtaining analytical expressions for the learning curve is usually not easy, and may even be impossible, in general, as the underlying equations are non-linear. We therefore limit our investigation to a restricted invasion-game, 
where for each percept $1\!\leq \!s\!\leq \!S$ there exist a single, unique, rewarded action, $a_s$, such that $h(s,a_s)$ is the only rewarded edge for this percept. In addition, we assume that percepts are shown one-by-one in a fixed order, starting with a randomly chosen percept at time $t_0$.
Let us denote with $t_0$ the (random) time between $0$ and $S-1$ when the specific percept $s=1$ is shown for the first time.
Fig.~\ref{fig:clipNetworkForAnalytics} illustrates the clip network as it develops for such a scenario.
Here, percept-clips and action-clips are shown in the first and second rows, respectively, and thick edges denote the rewarded edges. It is seen that the resulting configuration is highly structured (yet at each time step the strengths of the rewarded edges slightly vary, according to the order in which the percept are shown). This allows us to treat all percepts in the same manner, thereby simplifying our analysis.

First, we find an expression for the asymptotic value of the learning curve.
For a single agent, the blocking efficiency at time $t$, which we denote by $r^{(t)}$, is given by
\be
r^{(t)} \! = \!\!\sum_{s=1}^{S} \!\! P^{(t)}(s) p^{(t)}(a_s|s) \quad.
\ee
where $P^{(t)}(s)$ is the probability that the attacker shows the $s$ symbol (thereby exciting percept \tc{$s$}) and $p^{(t)}(a_s|s)$ is the hopping probability from percept $s$ to the rewarded action $a_s$ as defined in section \ref{sec:PS_formalism}. In our restricted invasion game, where percepts are shown in a fixed order (``regular training''), we have $P^{(t)}(s) = \delta((s - t \; \bmod \; S) + t_0)$. For large times, $t\to \infty$, the agent then evolves into a cyclically steady state described by $p^{(t)}(a_s|s) = p^{(t+S)}(a_s|s)$ and a corresponding blocking efficiency
\be
\nn r^{(t)}  \!\! & = & \!\! p^{(t)}(a_{s_0}|s_0) \\
& = & \!\!\! \frac{h^{(t)}(s_0,a_{s_0})}{h^{(t)}(s_0,\!a_{s_0})\!\! + \!\! \nn \sum_{a \neq a_{s_0}}^A \!\!\!\! h^{(t)}(s_0,\!a)} \\
&=& \!\!\! \frac{h}{h + (A-1)} \quad,
\label{eq:rt}
\ee
where $s_0$ refers to the percept presented to the agent at time $t$ and $a_{s_0}$ to the corresponding rewarded action. Note that $s_0$ and $a_{s_0}$ depend on the random time variable $t_0$ described above. In the second equality, we denote the $h$-value of the corresponding rewarded edge simply by $h$ and take into account that all unrewarded edges are of weight 1.

In the following, we will consider an ensemble of many agents, each of which is trained by the same percept history but may develop a different clip network. We are interested in the \emph{averaged} blocking efficiency, i.e.\ $\overline{r}$, and its asymptotic value, where the average is taken over the ensemble of agents. From (\ref{eq:rt}) we obtain the corresponding expression
\be
\overline{r} = \overline{\frac{h}{h + (A-1)}} \simeq \frac{\overline{h}}{\overline{h} + (A-1)}.
\label{eq:asymEfficiency}
\ee 
In the second step, we have made the essential approximation, which only holds if the distribution of the $h$-values is sufficiently narrow.
This requires that the weight of an edge does not change too much between subsequent rewards. In particular,
the damping should not change the $h$-value of an edge significantly while the $S-1$ other percepts are shown, i.e.\ $h^{(t_0+S-1)} = (1-\gamma)^{S-1}h^{(t_0)} \simeq h^{(t_0)} $. To first order in $\gamma$, this regime is characterized by $S\gamma \ll 1$.

With the approximation made in Eq.\ (\ref{eq:asymEfficiency}), finding an expression for $\overline{r}$ boils down to finding an expression for $\overline{h}$, i.e.\ the value of the rewarded edge at the cyclic steady state, averaged over many agents.
To express $\overline{h}$ we first rewrite the update rule, given in Eq.\ \eqref{eq:updateRuleOrgI}-\eqref{eq:updateRuleOrgII} to get:
\be
h^{(t+1)} - 1 = (1-\gamma) (h^{(t)} - 1) + \lambda   \quad,
\ee
where the reward $\lambda$ is assigned only to edges that were traversed during the last random walk.
Importantly, we note that when averaging over many agents, the rewarded edge is updated on average with a reward of $\lambda \overline{p}(a_s|s) = \lambda \frac{\overline{h}}{\overline{h} + A-1}$, that is by the reward times the probability to take the correct action.
The average of a rewarded edge, of a certain percept $s$, is therefore updated according to
\be
\overline{h}^{(t+1)} - 1 = (1-\gamma) (\overline{h}^{(t)} - 1) + \lambda  \frac{\overline{h}^{(t)}}{\overline{h}^{(t)} + A-1} \quad,
\ee
whenever the corresponding percept $s$ is encountered.
We recall that in our restricted invasion game, the percepts are shown one by one, in a fixed order, such that each percept $s$ is shown exactly once every $S$ time steps. The reward is therefore applied every $S$ time steps, and at the same time, the mere damping is applied $S-1$ times in between. In the asymptotic limit, the average value $\overline{h}$ thus reaches a (cyclically) steady state, where the average reward compensates the damping terms. This leads to the following recursion relations:
\be
\nn \overline{h}^{(t+1)}-1&=&(1-\gamma) (\overline{h}^{(t)} - 1) + \lambda \frac{\overline{h}^{(t)}}{\overline{h}^{(t)} + A-1} \\
\nn \overline{h}^{(t+2)}-1&=&(1-\gamma) (\overline{h}^{(t+1)} - 1) \\
\nn \overline{h}^{(t+3)}-1&=&(1-\gamma) (\overline{h}^{(t+2)} - 1) \\
\nn & \vdots &\\
\overline{h}^{(t+S)}-1&=&(1-\gamma) (\overline{h}^{(t+S-1)} - 1) \quad.
\label{recurrenceEq}
\ee
For the steady state we have $\overline{h}^{(t)} = \overline{h}^{(t+S)} \equiv \overline{h}$, which finally leads to (shown also in the original proposal \cite{BriegelCuevas12}):
\be
\overline{h}-1=(1-\gamma)^S(\overline{h}-1)+(1-\gamma)^{S-1} \lambda \frac{\overline{h}}{\overline{h}-1+A},
\label{eq:asymH}
\ee
a quadratic equation that can be solved for $\overline{h}$. Note that, due to the ordered percept excitation, there is still a cyclic time dependence in $\overline{h}$. For a given edge, the $\overline{h}$-value obtained from (\ref{eq:asymH}) refers to the times when a percept $s$ is shown that connects to this edge. 

\begin{figure}[htb]
	\begin{center}
		\begin{minipage}{9cm}
				\includegraphics[width=8cm]{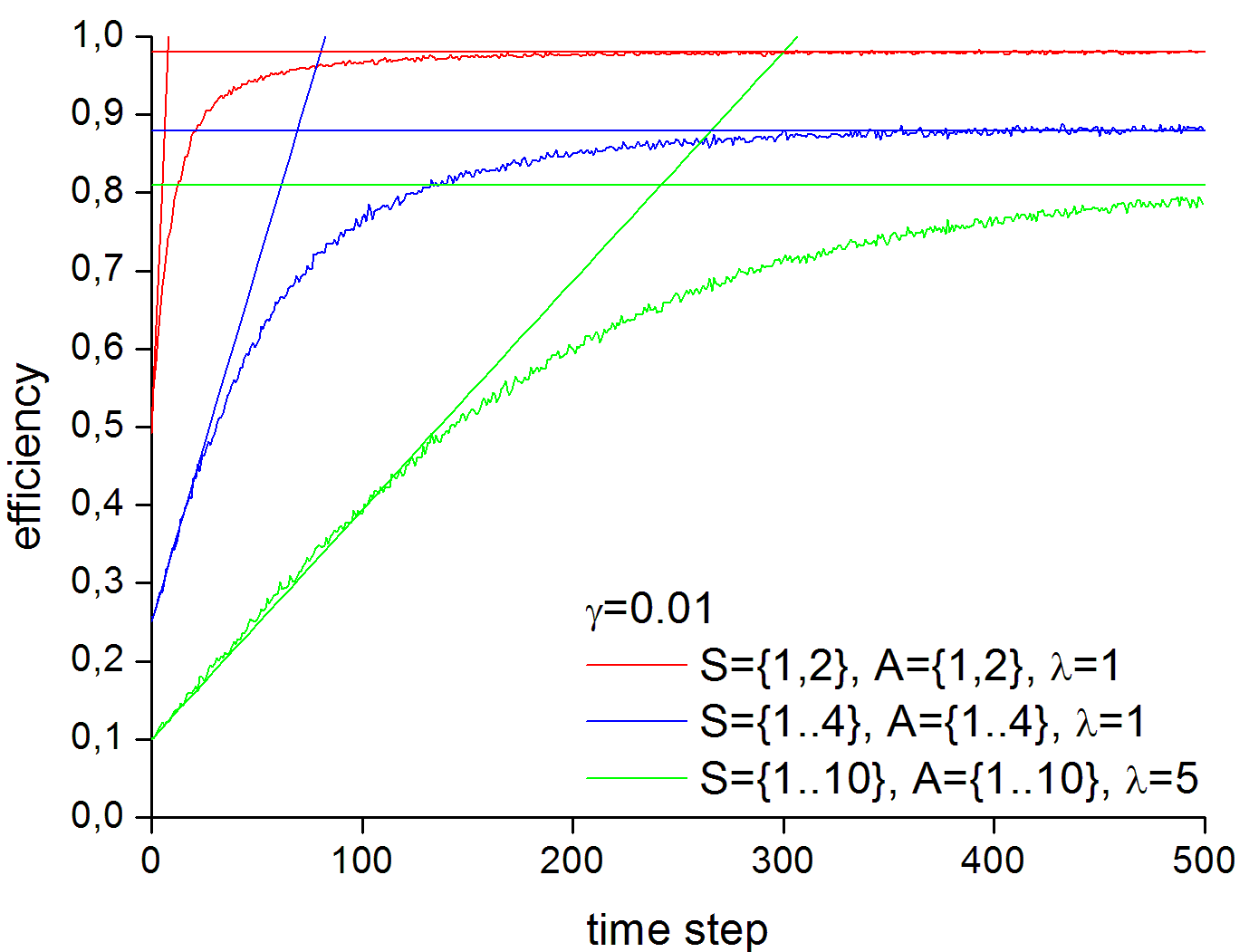}
			\end{minipage}
		\end{center}
	\caption{Numerical learning curves, obtained for a regular, \emph{unrestricted} invasion game,
accompanied with our analytical approximations of both the asymptotic efficiency, given in Eq.\ \eqref{eq:asymH}, and the curve's initial slope, given in Eq.\ \eqref{eq:slopeEfficiency}, as obtained for the \emph{restricted} invasion game (see text). Different values of $A=S$ and $\lambda$ are used, whereas $\gamma=0.01$ is fixed. The plot is taken from the original proposal \cite{BriegelCuevas12}.}
	\label{analyticComparison}
\end{figure}

In Fig.~\ref{analyticComparison} we show a comparison between the numerical learning curves for the invasion game (as described in section \ref{toyModel} and Fig.~\ref{fig:singleTrajectory}) and the approximate asymptotic value as computed from Eq.\ (\ref{eq:asymH}). It can be seen that the predicted asymptotic efficiencies are quite accurate, even though the numerical curves refer to a game with random percept training while the analytic values are obtained within our simplified scenario of regular percept training. 

We now turn to the study of the initial slope of the PS learning curve.
To that end we take into account the initial $h$-value, $h^{(0)}=1$, and estimate the derivative of the efficiency curve at $t=0$. Formally, we
obtain from (\ref{eq:asymEfficiency}) the expression
\be
\nn
\left. \frac{d\overline{r}}{dt}\right|_{t=0} &\!\simeq\!&  \frac{d}{d\overline{h}} \!\left. \Big( \frac{\overline{h}}{A-1+\overline{h}} \Big)\right|_{\overline{h}=1} \left. \frac{d\overline{h}}{dt} \right|_{t=0} \!\! \\ &=& \!\!\frac{A-1}{A^2} \frac{d\overline{h}}{dt} \!\simeq\!  \frac{A-1}{A^2} \frac{\Delta\overline{h}}{\Delta t}
\label{eq:derivativePeff}
\ee

To express $\Delta\overline{h}$ we look at the increase of the average $h$-value of a rewarded edge over $S$ time steps.
As before, we consider an ordered percept stimulation, meaning we hit one percept after the other in sequence.

To begin with, we focus on a specific (but arbitrary) rewarded edge.
We assume, as mentioned earlier, that the percept which connects to this edge is shown for the first time at $t_0$, somewhere from $0$ to $S-1$.
This leads, on average, to an update of the $h$-value of this percept-action edge given by:
$\overline{h}^{(t_0+1)} = 1+\frac{\lambda}{A}$.
During the subsequent $S-1$ steps, different percepts will be shown, resulting in $S-1$ damping steps for our particular $h$-value.
Following the same set of equations as given in \eqref{recurrenceEq}, together with the initial condition $\overline{h}^{(t_0)}=\overline{h}^{(0)}=1$, we arrive at:
\be
\Delta \overline{h}\!=\!\overline{h}^{(t_0+S)}-\overline{h}^{(t_0)}\!=\!\overline{h}^{(t_0+S)}-\overline{h}^{(0)}\!=\!\frac{\lambda (1\!-\!\gamma)^{S-1}}{A}.
\label{eq:deltaHslope}
\ee
This describes the increase of the $h$ value over a cycle, starting at time $t_0$ and ending at time $t_0+S$. Consider now the time $t_0$ as a random variable, distributed equally between $0$ and $S-1$, corresponding to an ensemble of (regularly trained) agents with different initial percepts. The initial ``slope'' of the learning curve, as estimated by 
$\frac{\Delta \overline{h}}{\Delta t} = \frac{\Delta \overline{h}}{t_0+S}$, will depend on $t_0$ and thus be different for the different sub-ensembles. Averaging over $t_0$ gives rise to an effective $\frac{1}{\Delta t} = \frac{1}{S}\sum_{t_0=0}^{S-1}\frac{1}{t_0+S}$, which can be 
approximated by a more compact (but less accurate) expression $\Delta t \simeq S-1 + \frac{S}{2}$.
Together with \eqref{eq:deltaHslope} and \eqref{eq:derivativePeff} this provides the following heuristic approximation: 
\be
\frac{d\overline{r}}{dt} \simeq \frac{\lambda (1-\gamma)^{S-1}(A-1)}{A^3(S-1+\frac{S}{2})}
\label{eq:slopeEfficiency}
\ee

In Fig.~\ref{analyticComparison} we plot the lines
$y(x) = \frac{d\overline{r}}{dt}x + \frac{1}{A}$ for different parameters of $\lambda, \gamma, A$, and $S$.
Here we used the expression of $\frac{d\overline{r}}{dt}$, given in Eq. \eqref{eq:slopeEfficiency} as the slope, and took into account that the initial value of the efficiency is given by $1/A$.
It is seen that in all three cases, the resulting lines are approximately tangential to the numerical learning curves in the initial time steps.
The analytic approximations are a useful tool to predict the qualitative change in the agent's learning behavior when the different parameters are changed, and to check the plausibility of the numeric results.

\subsection{The damping parameter $\gamma$} \label{sec:damping}
The constant damping of the $h$-values can be interpreted as an ongoing ``forgetting" over time. This important feature of the PS model allows for a weakening of connections between clips, and thereby making it possible for the agent to adapt to changing environments. Using a positive $\gamma$ (in this work typical values are around $0.02$) is further motivated from the point of view of an embodied agent: physically, it assures finite $h$-values of the edges, and biologically, as said, it represents a natural forgetting over time.

We note, however, that the use of damping affects the overall performance of the PS agent. For example, we note that when $\gamma$ is positive the asymptotic efficiency of the PS agent is not optimal, as can be seen e.g.\ in Fig.~\ref{fig:errorBarsPlot} and Fig.~\ref{analyticComparison}. 
Without damping ($\gamma=0$), the asymptotic efficiency is equal to unity. This is a general property of PS, which will be encountered many times throughout this paper.

The effect of the damping is even more pronounced as the size of the ECM network is increased: the more edges there are, the more damping steps occur on average between two successive rewarding steps.
This is demonstrated in Fig.~\ref{analyticComparison}, where increasing the dimension $S$ and/or $A$ results in a decreased asymptotic efficiency. The negative effect of the damping can, however, be tackled in various ways: The simplest way would be to reduce the value of $\gamma$. When $\gamma$ is made smaller, the asymptotic value can be made arbitrarily high. Another possibility is to increase the reward $\lambda$. This directly increases the $h$-values, thereby compensating for the damping and increasing the efficiency. The third possibility is to increase the reflection time $R$ 
to boost the asymptotic efficiency, as shown below. 

Finally we note that, even though the use of a damping term is well motivated, it is not a necessary ingredient in the general scheme of projective simulation, which may be implemented with many different learning rules.

\subsection{Scaling of learning times\label{sec:scaling}}
In what follows we examine the scaling properties of PS learning times, when increasing
the number of possible percepts and/or actions.
We emphasize that while in section \ref{sec:approxSol} we studied the initial slope of the learning curve as a first indicator for the learning time, here we are interested in the actual learning time, which we define as:
\bd
\item [Learning time] The number of steps needed for the averaged efficiency to reach a certain threshold fraction of its asymptotic value, for the first time.
\ed
In the remainder of this paper, we set the threshold fraction to be 0.9.

We note that this definition of the learning time may allow situations in which the learning time turns out to be misleadingly short, merely because the asymptotic efficiency is very low (hence reaching the threshold by statistical fluctuations only). To avoid such situations, we either set $\gamma=0$ or rescale it properly, as explained in section \ref{sec:approxSol}. This way the asymptotic value is either equal to unity or sufficiently high ($>0.5$).

\begin{figure}[htb]
	\begin{center}
		\begin{minipage}{9cm}
				\includegraphics[width=8cm]{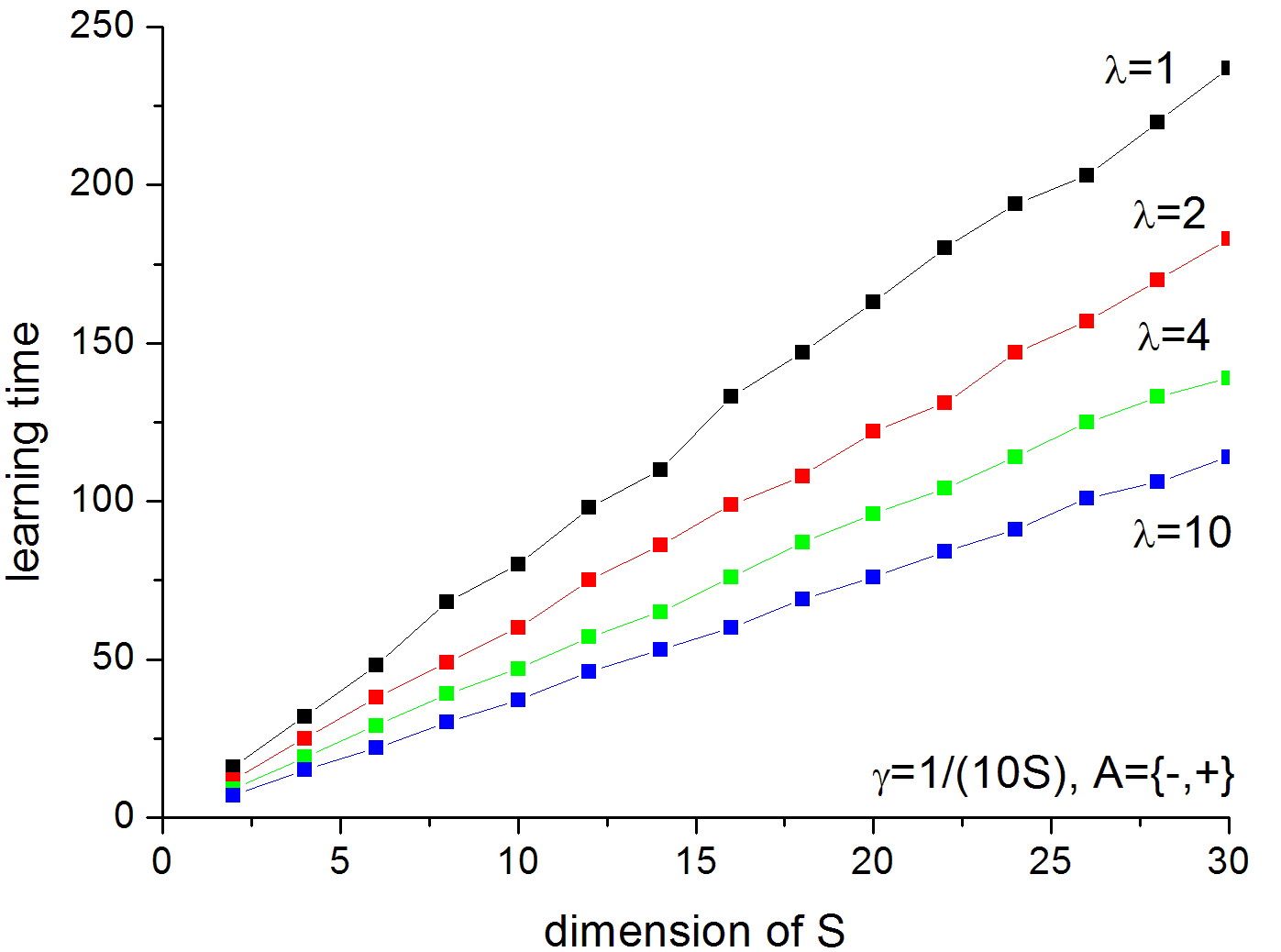}
			\end{minipage}
		\end{center}
	\caption{Learning time of PS shown as a function of the size $S$ of the percept space ${\cal S}$, for different values of the reward $\lambda$. It is seen that the PS learning time scales linearly with $S$, and that for larger values of $\lambda$ the learning time increases slower with $S$.}
	\label{fig:learningTimeS2_PS}
\end{figure}

We start with the scaling of the learning time when the number of possible percepts is increased. Fig.~\ref{fig:learningTimeS2_PS} shows the learning time as a function of the size $S$ of percept space, for different values of the reward $\lambda$. The  number of actions $A=2$ is kept fixed, and there is exactly one rewarded action per percept. It is seen that the learning time scales linearly with increasing $S$. This linear scaling arises because each percept has to be encountered the same number of times to allow the agent to learn the rewarded action.
It is further seen that increasing the reward reduces the learning time as hinted at by Eq.\ \eqref{eq:slopeEfficiency}. This is because fewer steps are needed before the higher steady state $h$-value is achieved. Here we set $\gamma=\frac{1}{10 S}$ to compensate for the decrease in asymptotic efficiency.  We remark that when compared to QL, we observe a similar trend of the learning time with respect to increasing the reward (not shown).

\begin{figure}[htb]
	\begin{center}
		\begin{minipage}{9cm}
				\includegraphics[width=8cm]{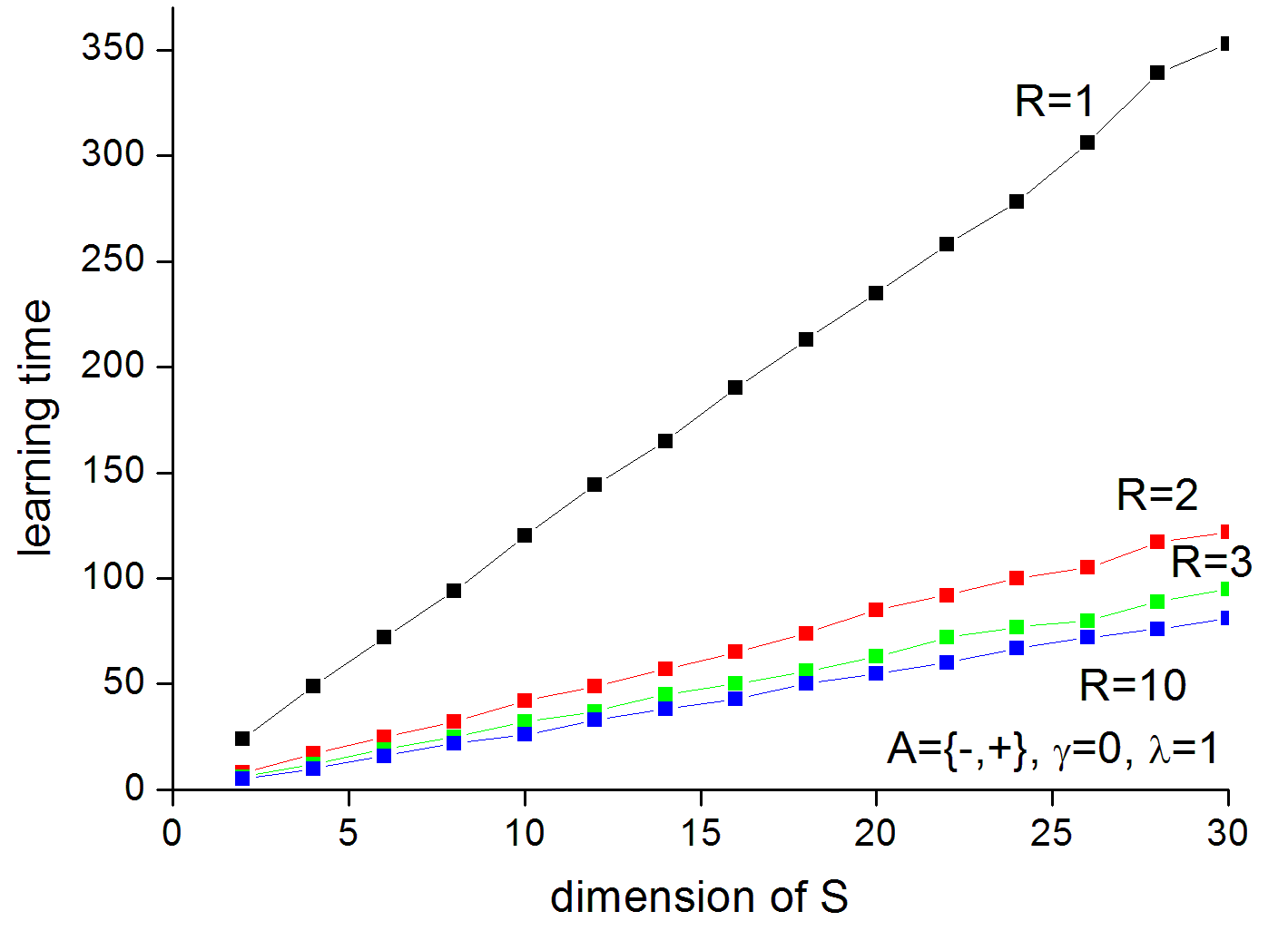}
			\end{minipage}
		\end{center}
	\caption{Learning times of PS shown as a function of the size of the percept space $S$, using different values of the reflection $R$. It is seen that the PS learning time scales linearly with $S$, and that for larger values of $R$ the learning time increases slower with $S$. The plot is reprinted from the original proposal \cite{BriegelCuevas12}, with small modifications.}
	\label{fig:learningTimeS_PS}
\end{figure}

In Fig.~\ref{fig:learningTimeS_PS} we examine the effect of changing the reflection number $R$ (see definition in section \ref{sec:PS_formalism}) on the learning time. We set $\gamma=0$, so no damping occurs. It is seen that
increasing the number of reflections does not effect the scaling behavior, which stays linear, but that it reduces the overall learning time, in a similar effect to an increase in the reward.
This beneficial behavior of using multiple reflection is very general within the PS scheme, as it (in most cases) boosts the entire performance in terms of both the asymptotic efficiency as well as the learning time of an agent.

\begin{figure}[htb]
	\begin{center}
		\begin{minipage}{9cm}
				\includegraphics[width=8cm]{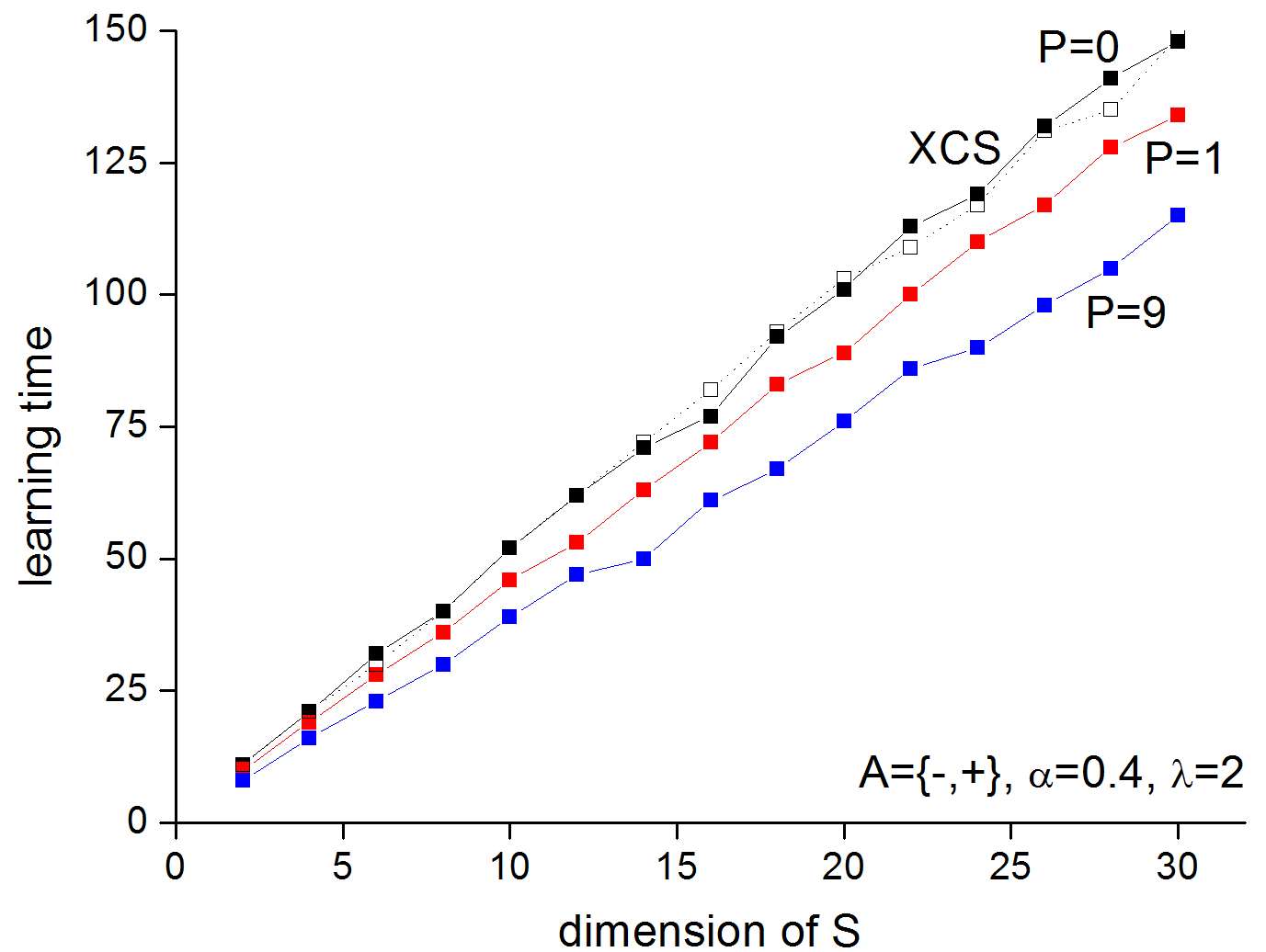}
			\end{minipage}
		\end{center}
	\caption
{Learning times of QL (with Dyna-style planning) shown as a function of the size of the percept space $S$, using different number of planning steps $P$. 
The reward is increased to $\lambda=2$ for achieving a reasonably high asymptotic efficiency while maintaining to use a softmax policy, and $\alpha$ is a learning rate parameter, defined in appendix \ref{appx_QL}. It is seen that the QL learning time scales linearly with S, and that for larger values of $P$ the learning time increases slower with $S$. A learning-time plot is also shown for XCS (depicted with hollow squares) using the same learning rate $\alpha$.}
	\label{learningTimeS_Q}
\end{figure}

Fig.~\ref{learningTimeS_Q} shows the learning time as a function of the size of  the input space, as performed by QL (accompanied with Dyna-style planning) and XCS.
It is seen that with these schemes too, the learning time scales linearly with the size of the input space. Moreover, it is shown that increasing the number of planning steps, $P$, for the Dyna-style planning scheme, reduces the learning time, with a similar effect as increasing the reflection parameter, $R$, in PS (one should note, however, that using more planning steps does not lead to an increase in the asymptotic efficiency, whereas using a higher reflection parameter, $R$, does \cite{BriegelCuevas12}).

Last, we studied the scaling of the PS learning time when both input space $S$ and action space $A$ are increased together, such that $S=A$.
We found that the learning time scales quadratically, implying a linear scaling when $A$ alone is increased.  A similar quadratic behavior was also observed for QL and XCS. We thus conclude that in such simple scenarios, all three models behave qualitatively in a similar way with respect to the scaling of the learning time when increasing the problem size (input and/or action dimensions).

\section{Temporal correlations\label{sec:tempCorr}}
In this section we investigate the performance of PS in situations that are far more complicated than those presented until now. In particular, we study scenarios in which the rewards the agent receives may depend not only on the particular action it takes at present, but also on actions it made previously, that is we use non-Markovian rewarding schemes. To such situations we refer as \emph{temporal correlations}.

\subsection{``Afterglow'' mechanism} \label{sec:afterglow}
The PS scheme, as it was introduced so far, has no efficient mechanism to handle situations of the form of ``temporal correlations''. It does, however, provide a flexible paradigm that can be naturally extended. Here, we
generalize the \emph{excitation} process, i.e.\ the dynamics of excitation propagation in the ECM, to account for temporally correlated scenarios.
It is expedient to associate a certain state of ``excitation'' also with edges that have been used during the random walk. We refer to such an excitation informally as ``edge glow''. 
An excited or glowing edge indicates that this edge should be strengthened if the subsequent action is rewarded.
In what follows, we allow edge excitations to decay slowly, step by step, instead of fully decaying after a single step only. This means that an 
edge that was not used during the latest random walk may nevertheless be still (partially) excited, because it was visited in previous time steps.
The edges are then strengthened in straight correlation to the level of their excitation: the stronger the excitation, the larger is the reward. 
As before, edges that are not excited at all get no reward. We refer to this slow decay of edge excitation as \emph{afterglow}.

Formally, the afterglow is implemented using a new degree of freedom: a parameter $g$ called ``glow'', which is attached to each of the edges in the clip network.
Initially, $g$ is set to zero for all edges. Then, once an edge is visited during the random walk, its glow parameter is set to $g=1$. In subsequent steps $g$ is damped at each time step toward zero with rate $\eta$, according to:
\be \label{eq:glowDamping}
g^{(t+1)}=g^{(t)}-\eta g^{(t)}, \quad   0 \le \eta \le 1,
\ee
and the update rule of Eq.\ \eqref{eq:updateRuleOrgI}-\eqref{eq:updateRuleOrgII} is modified to:
\be \label{eq:updateEdgeRule}
\nn h^{(t+1)}(c_1, c_2) &=& h^{(t)}(c_1, c_2) - \gamma (h^{(t)}(c_1, c_2)-1) \\
&\ & + \lambda g^{(t)}(c_1,c_2),
\ee
where $g^{(t)}(c_1,c_2)$ stands for the $g$ value of the edge connecting clip $c_1$ to clip $c_2$ at time $t$.
As before, only excited edges may be strengthened when a reward $\lambda$ is given. The difference is that the edges may now be partially excited, allowing for a partial reward, according to the strength of their excitation. 
Therefore, an unrewarded edge (i.e.\ an edge associated with an unrewarded transition) may eventually be rewarded if subsequent actions are sufficiently rewarded, thus enabling a non-greedy choice of actions. 
We note that $\eta=1$ amounts to no ``afterglow''. Accordingly, all plots shown previously can be obtained within the afterglow scheme, by simply setting $\eta=1$.

Finally, we remark that the use of reflection ($R > 1$, see section \ref{sec:PS_formalism}) may sometimes conflict with the afterglow scheme. This is because reflection leads to a pseudo-greedy strategy: with it the agent is inhibited to take any action that was unrewarded in the previous step, even if it is more beneficial in the long run. No unrewarded paths can thus be explored, which would limit the effect of afterglow. In what follows we set $R=1$, i.e.\ no reflection is used.

\subsection{The $n$-ship game}
To examine the utility of afterglow we employ the \emph{$n$-ship game}, a variant of the invasion game. It consists of a single door and several attacking ships. The defender-agent can take one of two actions: to block and to not-block. The attacking ships arrive in sequence and for each of them the agent has to decide whether it should block it or not. Temporal correlations are introduced to the game by using a time-dependent rewarding scheme. This means that an action is rewarded/not-rewarded according to both present actions and actions taken at previous time steps.

In the simplest version of the $n$-ship game, only two ships arrive, i.e.\ $n=2$. A temporal correlation is then implemented through the following rewarding scheme:
if the agent blocks the first ship, it gets a small reward $\lambda_S$, but will not get any reward for neither blocking nor not-blocking the second. On the other hand, if it does not block the first ship, it will get a larger reward
$ \lambda_L > \lambda_S $ for blocking the second.
Thus the agent has to learn to let the first ship pass, despite being rewarded for blocking it, and aim at blocking only the second ship.

\begin{figure}[h]
	\begin{center}
		\begin{minipage}{9cm}
				\includegraphics[width=8cm]{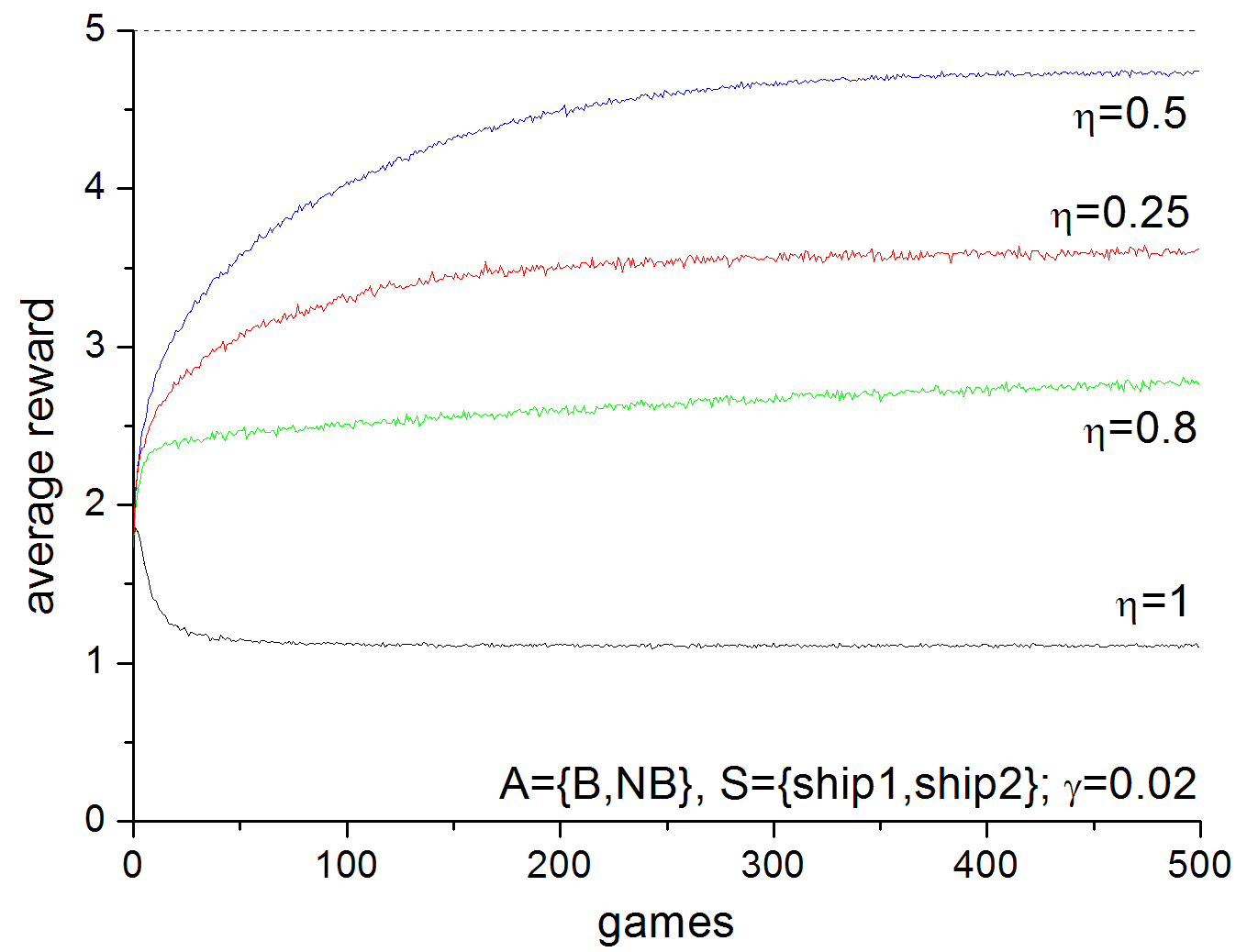}
			\end{minipage}
		\end{center}
	\caption{Averaged reward of PS in the 2-ship game. A reward of 1 is given for blocking the first ship and a reward of 5 is given for blocking the second, \emph{if} the first ship was not blocked before (see text).
The maximal achievable reward of 5 cannot be reached due to $\gamma > 0$. For an equiprobable choice of both actions the average reward is~$7/4$.}
	\label{fig:shipGame2_PS}
\end{figure}

Fig.~\ref{fig:shipGame2_PS} shows the performance of the PS agent in the simplest 2-ship game.
Here, each game is composed of two time steps and the games follow each other continuously. We emphasize that from the point of view of the agent there is no notion of a distinct game, so that all it perceives is a continuing process with ships number one and two following one another. The sum of rewards given for both ships is plotted as a function of the number of games.
One can see that without afterglow ($\eta=1$) the agent cannot reach the high reward, because most of the time it greedily blocks the first ship. However, when $\eta$ is decreased, the afterglow is turned on, and a higher average reward is achieved. This indicates that using the afterglow is indeed beneficial for temporally correlated scenarios.

Fig.~\ref{fig:shipGame2_PS} further indicates that below a certain value of $\eta$, the performance of the agent deteriorates. This can be seen by comparing $\eta=0.5$ (in blue) to $\eta=0.25$ (in red).
The success of afterglow thus depends on the $\eta$ parameter (see
Eq.~\eqref{eq:glowDamping}): when $\eta$ is too large no memory of edge excitations is carried forward to successive steps, whereas when it is too small the excitations are never damped. In the later case, excitations from all previous steps persist (including  previous games!), leading to an undesirable circumstance of over memorizing and rewarding all clip transitions taken in the past, even though they may be unrelated, resembling a state of ``confusion".

\begin{figure}[h]
	\begin{center}
		\begin{minipage}{10cm}
				\includegraphics[width=8cm]{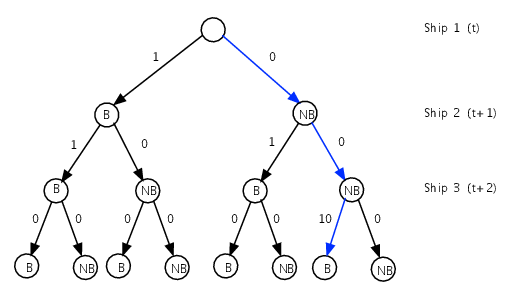}
			\end{minipage}
		\end{center}
	\caption{Reward scheme for the n-ship game with $n$=3. Here, B and NB denote blocking and non-blocking actions, respectively, and the numbers along the arrows denote the corresponding rewards. The blue sequence of actions denotes the optimal strategy.}
	\label{fig:nshipGame}
\end{figure}

We next consider a general $n$-ship game. Here the environment rewards the blocking of any of the first $n-1$ ships with a small reward of $\lambda_S$, but rewards the blocking of the last ship with a large reward if and only if all previous ships were let pass. Fig.~\ref{fig:nshipGame} illustrates this rewarding scheme for $n=3$. It can be seen that the number of possible paths grows exponentially with $n$, which makes the problem very hard. We remark that in the $n$-ship game we rescale the large reward $\lambda_{max}=(n-1)\cdot\lambda_L$ so that it remains beneficial to block only the last ship. In what follows we use $\lambda_S=1$ and $\lambda_L=5$.

\begin{figure}[h]
	\begin{center}
		\begin{minipage}{9cm}				\includegraphics[width=8cm]{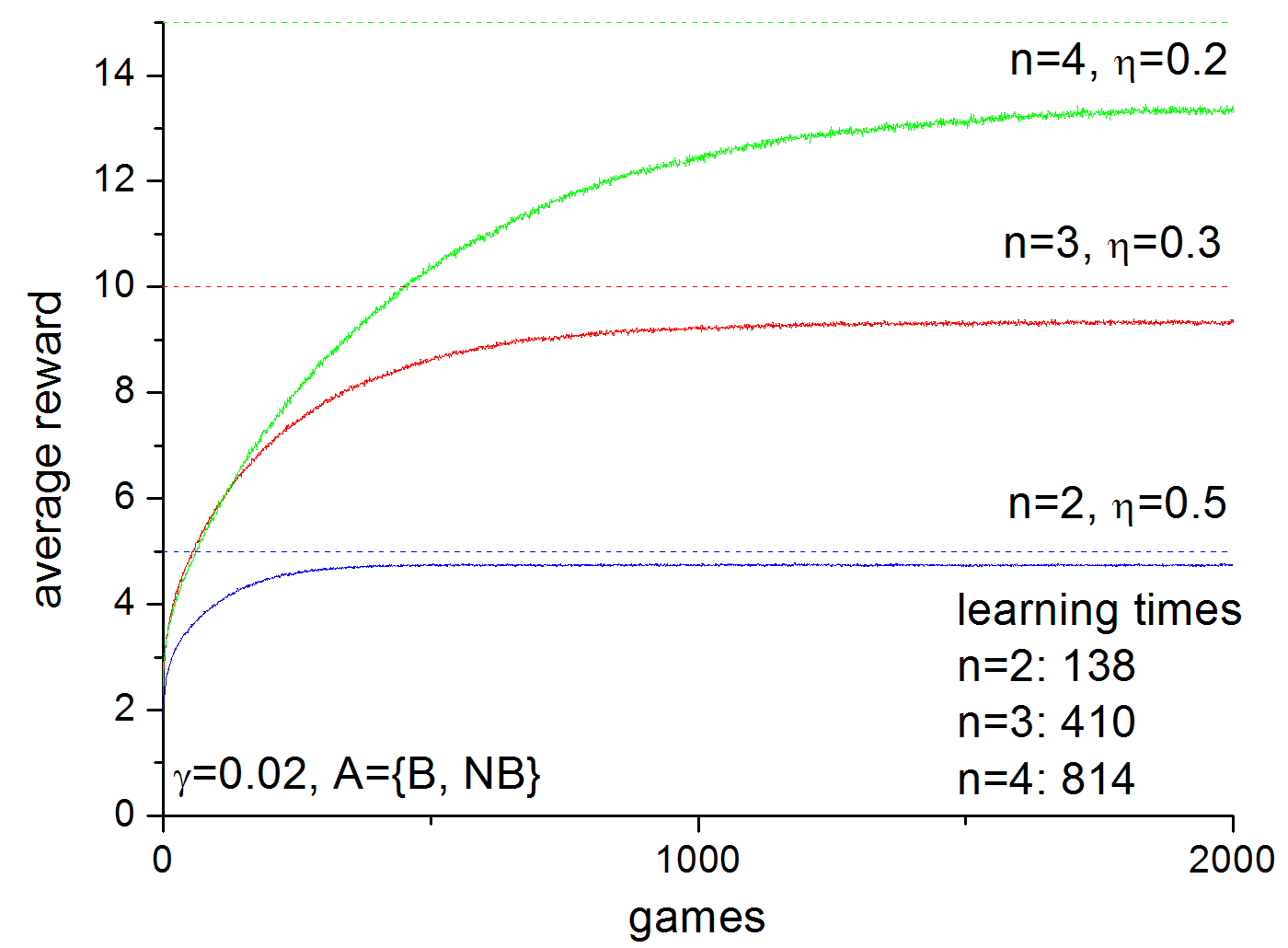}
			\end{minipage}
		\end{center}
	\caption{Averaged reward of PS in the $n$-ship game for $n=$2, 3, and 4 shown in solid lines. Dashed lines indicate the maximum achievable reward in a single game. Learning times for each $n$ are also indicated. For each curve an (nearly) optimal $\eta$ was used. For comparison: A greedy strategy results with a maximal achievable reward of $n-1$, and the average reward of an untrained agent of equiprobable choice of both actions is ~$1.75, \; 2.25$ and $2.44$ for $n=2,3,4$, respectively.}
	\label{fig:efficiencyN-layerShipGame_PS}
\end{figure}

Fig.~\ref{fig:efficiencyN-layerShipGame_PS} shows in solid lines the performance of PS in the $n$-ship game for $n=2,3,4$, accompanied with dashed lines at 5, 10, and 15, respectively, to indicate the maximal achievable reward for each $n$. As before, each game is composed of $n$ time steps, one for each ship. The sum of all rewards given by the $n$ ships is then plotted as a function of the number of games. For each curve, an optimal $\eta$ parameter was used. It is seen that for each of these curves the asymptotic reward is higher than $n-1$ which is the maximal achievable reward when a naive greedy strategy is used (and much higher than the averaged reward achieved by an untrained agent  with equiprobable action choice, see caption).
This indicates that most of the PS agents successfully adopt a non-greedy strategy, even when $n$ increases.

Fig.~\ref{fig:efficiencyN-layerShipGame_PS} further indicates the learning time for each $n$. It is seen that the learning time roughly doubles for each additional layer. Indeed the learning time of PS scales exponentially with $n$ (not shown). This is however to be expected, as the problem of merely finding the most rewarded series of actions becomes exponentially hard as $n$ increases (see Fig.~\ref{fig:nshipGame}).

\begin{figure}[h]
	\begin{center}
		\begin{minipage}{4cm}
				\includegraphics[width=4cm]{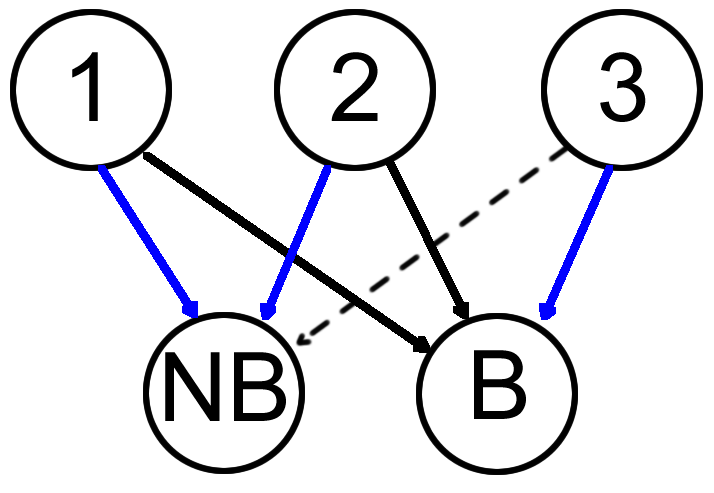}
			\end{minipage}
		\end{center}
	\caption{A schematic illustration of the PS clip network, as built during learning for the 3-ship game. The dashed edge is the weakest, solid black edges are stronger, and blue edges are the strongest.}
	\label{fig:shipGameClip_PS}
\end{figure}

The success of PS to learn in temporally correlated scenarios owes itself to the afterglow mechanism, which introduces an implicit notion of time to the agent's memory (with no need of an explicit ``time counter"). To illustrate better the underlying process, we show in Fig.~\ref{fig:shipGameClip_PS} a schematic drawing of the clip network, as it is built up during many 3-ship games (see also Fig.~\ref{fig:nshipGame}).
Percept clips (\ding{172},\ding{173},\ding{174}), shown on top, indicate the ships numbers, whereas action clips (\textbf{NB}, \textbf{B}), shown at the bottom, indicate non-blocking, and blocking actions, respectively.
The number of clips is thus given by $n+2$: at each round only a single percept is encountered and the agent is not supplied with the history of its previous actions. A dashed black arrow marks the weakest edge, which is never rewarded. Solid black edges are stronger (not necessarily equal). They are always rewarded, albeit with small reward $\lambda_S$. Last, blue edges are the strongest (not necessarily equal), thereby effecting the agent the most. It is seen that for the first two ships the edge to the non-blocking action is stronger than the edge to the blocking action, even though the blocking action is the rewarded one. This is desirable and achieved via strengthening the non-blocking edges \emph{indirectly} after blocking the third ship and getting a large reward. It is thus the strength of these two blue edges, whose emergence is not trivial and allows the agent to take non-greedy actions that are more beneficial in the long run.

The afterglow scheme is however not optimal. Fig.~\ref{fig:efficiencyN-layerShipGame_PS} also indicates that the performance of PS deteriorates as $n$ increases: more and more agents fail to learn to avoid the greedy actions, as reflected in the reduced asymptotic values of the achievable rewards, relative to the maximal ones. This is because the time length of correlations between different actions becomes longer and longer, making it difficult for the PS to construct and maintain an optimal clip network. Moreover, when temporal correlations extend along many steps, the excitation should persist longer before decaying. Hence, we expect an inverse relation between the optimal $\eta$ value and the length of the temporal correlation.

\begin{figure}[hb]
	\begin{center}
		\begin{minipage}{9cm}
				\includegraphics[width=8cm]{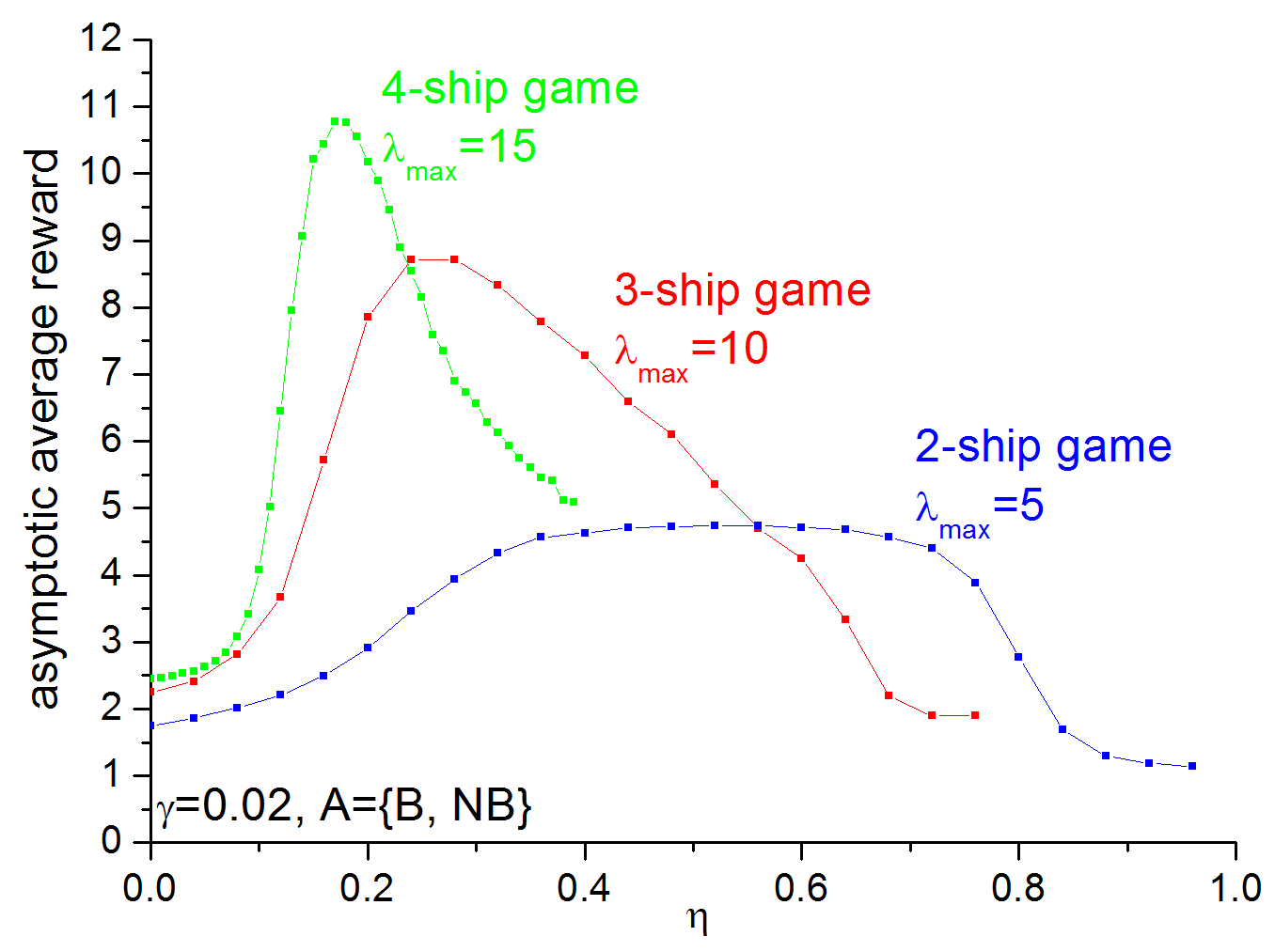}
			\end{minipage}
		\end{center}
	\caption{Achievable asymptotic reward shown as a function of $\eta$ for the $n$-ship game with $n=$ 2, 3, and 4. For each $n$ an optimum value of $\eta$ is depicted, for which the agent remembers enough, and yet not too much.
It can be observed that the peaks become narrower as $n$ increases, that is for increasing correlation times.}
	\label{etaResonance}
\end{figure}

To understand the dependence over the $\eta$ parameter in a quantitative manner, we plotted the asymptotic averaged reward of the PS agent as a function of the $\eta$ parameter, for the 2-,3-, and 4-ship game, as shown in Fig.~\ref{etaResonance}. We see that indeed an optimal $\eta$ value exists. It is further seen that each $n$-ship game, has a different optimal $\eta$, which decreases when $n$ is increased, as expected.

For comparison, we turn now to study the performance of QL and XCS in temporally correlated scenarios (see Appendices ~\ref{appx_QL} and ~\ref{appx_LCS}, respectively). The QL scheme relies on a temporal-difference mechanism~\cite{SuttonBarto98} which allows it to propagate rewards to earlier time steps. This makes QL a natural reference scheme to begin with.
We therefore implemented a QL agent and let it play the $n$-ship game.
To that end we used a Q-function that has $2n$ state-action entries, i.e.\, for each time step the corresponding ship number is given as a state of the environment, for which the agent may take one of two actions. The size of the Q-function thus grows linearly with $n$ in this setup.
Further, we have chosen QL parameters that are (nearly) optimal for that problem, given by a learning rate of $\alpha=0.4$ and a discount factor of $\gamma=1$ (for parameter definitions see Appendix \ref{appx_QL}).

\begin{figure}[h]
	\begin{center}
		\begin{minipage}{9cm}
				\includegraphics[width=8cm]{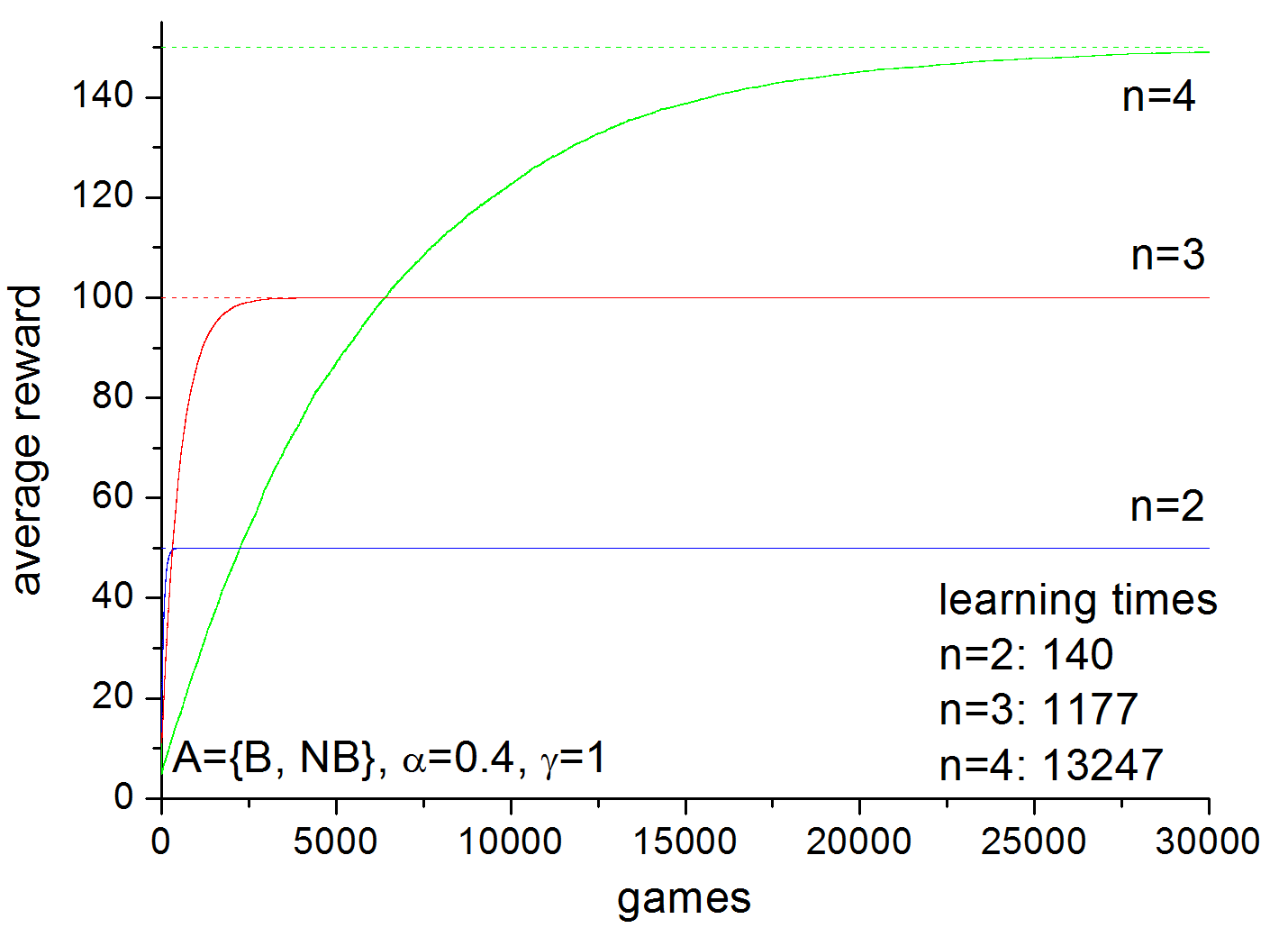}
			\end{minipage}
		\end{center}
	\caption{Average reward of QL in the $n$-ship game for $n=$2, 3, and 4 shown in solid lines. Dashed lines indicate the maximum achievable reward in a single game. Learning times for each $n$ are also indicated. For comparison: A greedy strategy results with a maximal achievable reward of $n-1$, and the average reward of an untrained agent of equiprobable choice of both actions is ~$13, \; 13.5$, and $10.87$ for $n=2,3,4$, respectively.}
	\label{fig:shipGameAvgRewardLinearStates_QL}
\end{figure}

We found that the QL agent does not perform well in the $n$-ship game. In particular, the scaling of the maximal reward $\lambda_{max} = (n-1)\cdot\lambda_L$ is not sufficient and the QL agent clings to the less beneficial greedy strategy. This, however, can be overcome by rescaling the maximal reward $\lambda_{max}$ even further. Fig.~\ref{fig:shipGameAvgRewardLinearStates_QL} shows the averaged reward obtained by QL for the $n$-ship game with $n\!=\!2,3,4$ where the maximal reward is scaled as $\lambda_{max,QL}=10\!\cdot\!(n-1)\!\cdot\!\lambda_L$. It is seen that the QL performance resembles that of PS as shown in Fig.~\ref{fig:shipGame2_PS}.
QL differs, however, by achieving a nearly optimal performance, yet at the cost of very large learning times. The scaling of the learning time in QL is exponential too (not shown), by an approximated factor of ten for each additional ship (see learning times for $n\!=\!2,3,4$ in Fig.~\ref{fig:shipGameAvgRewardLinearStates_QL}).

XCS shares many parallels with QL regarding the performance in the $n$-ship game. The learning curves are qualitatively similar to those of QL for the same reward scheme and learning parameters (not shown). For our implementation we have used a population of $2n$ classifiers, and disabled the genetic algorithm as it only degraded performance. Similarly to QL, the XCS requires the scaling up of the reward to learn the non-greedy strategy.

Before concluding, we remark that the performance of all three models can of course be significantly improved by providing the entire history of previous actions to the agent, i.e.\ by having an explicit representation of the past in the agent's memory. In particular, one can associate each possible path of states and actions (see Fig.~\ref{fig:nshipGame}) with a percept, a state, or a classifier. However, this approach, which is exponentially expensive in space for PS and QL, is against the spirit of an embodied (and thus finite) agent architecture and trivializes the problem in a way we want to avoid. 

We conclude that all three models perform qualitatively the same in the $n$-ship problem.
Specifically, in terms of space and time requirements they all scale similarly: linear with space and exponential with time. Quantitatively, both QL and XCS achieve an optimal efficiency but are relatively slow. In addition, the reward must be significantly scaled upward to allow for meaningful learning.
PS, on the other hand, cannot achieve a fully optimal efficiency, due to forgetting, but learns very fast due to the afterglow mechanism, as explained above.

\subsection{The $2$-ship game with $A$ actions}
To add a further complication to the $2$-ship game, we now allow for the more general case of $A$ different actions (instead of just two), whenever a ship arrives.
Against the first ship, all actions are rewarded with a small reward $\lambda_S=1$, except of a single action $a_{0}$ which is not rewarded at all. Then for the second ship, half of the actions are never rewarded, whereas half of the actions are highly rewarded with $\lambda_L=15$, but only if the unrewarded action $a_{0}$ was previously taken (otherwise, they are not rewarded either). The idea, as before, is that in order to maximize the long-term reward, the agent must avoid making any of the rewarded actions in the first round, to eventually obtain a much larger reward in the second round. The agent, however, has now many optional actions at its disposal, making it much more difficult and unlikely to make the smart choice per chance and act with the single unrewarded action $a_{0}$ against the first ship.

\begin{figure}[t]
	\begin{center}
		\begin{minipage}{7cm} \includegraphics[width=\linewidth]{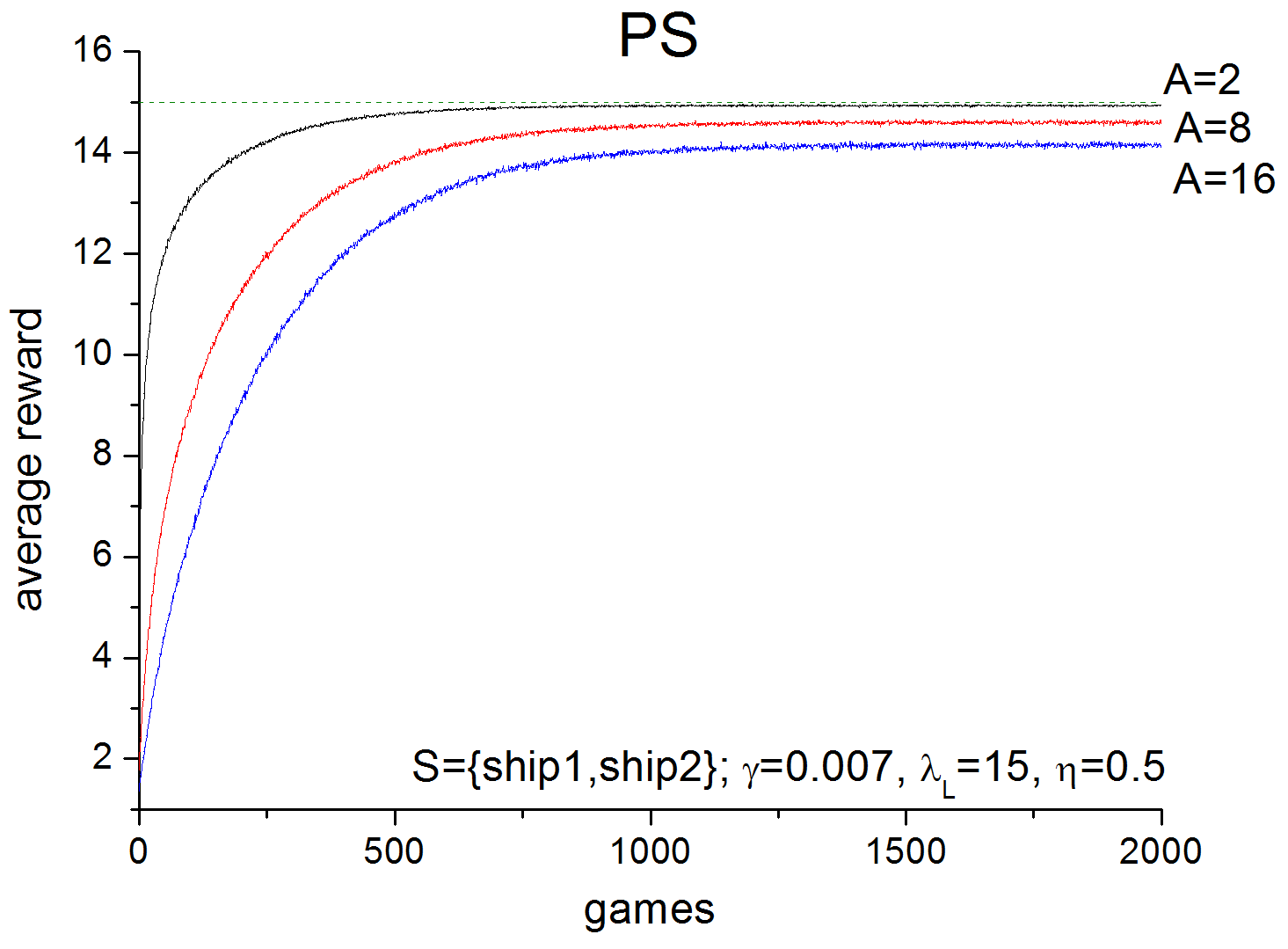}
			\end{minipage}
		\end{center}
	\begin{center}
		\begin{minipage}{7cm}
				\includegraphics[width=\linewidth]{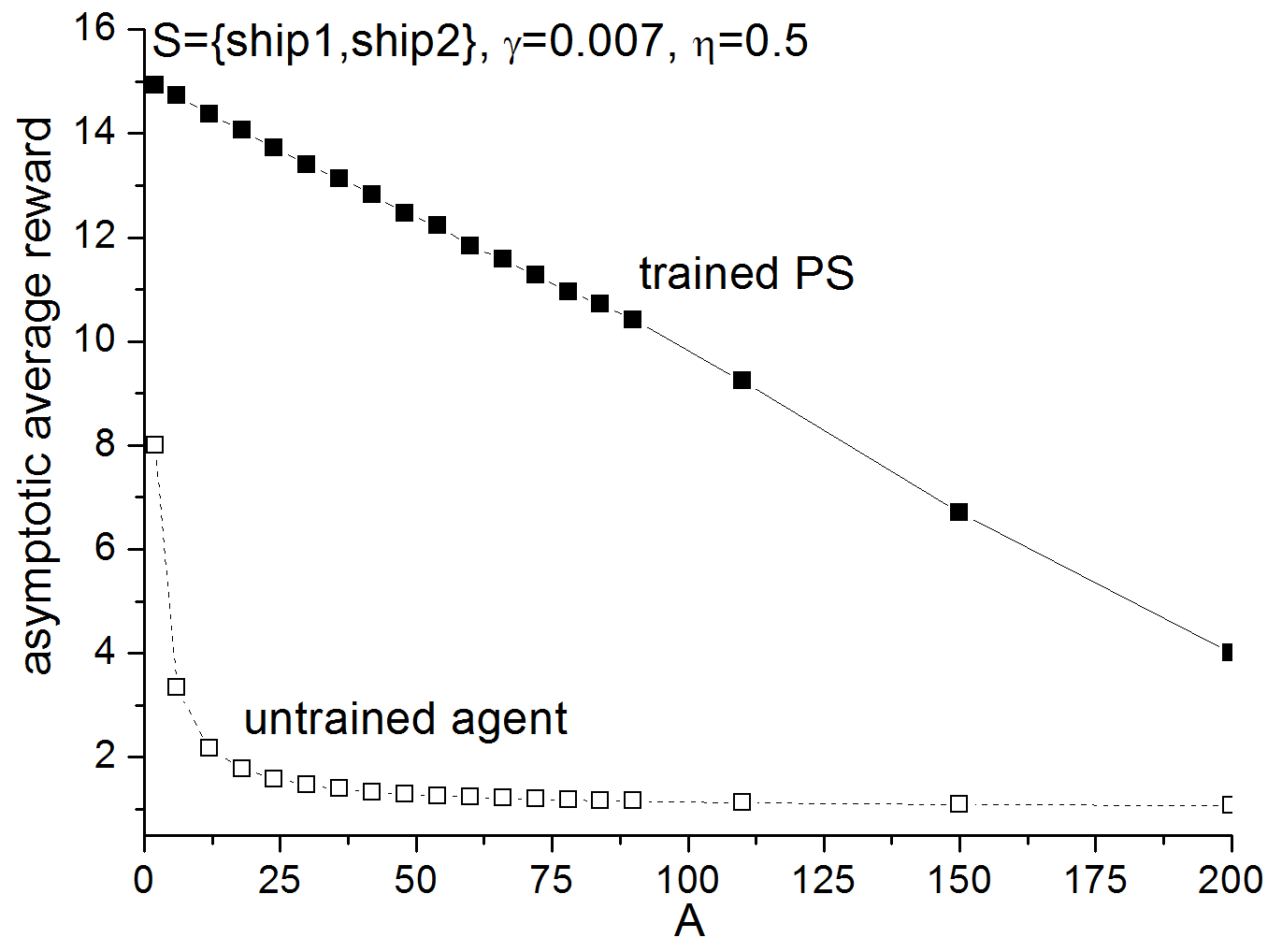}
			\end{minipage}
		\end{center}
	\begin{center}
		\begin{minipage}{7cm}
				\includegraphics[width=\linewidth]{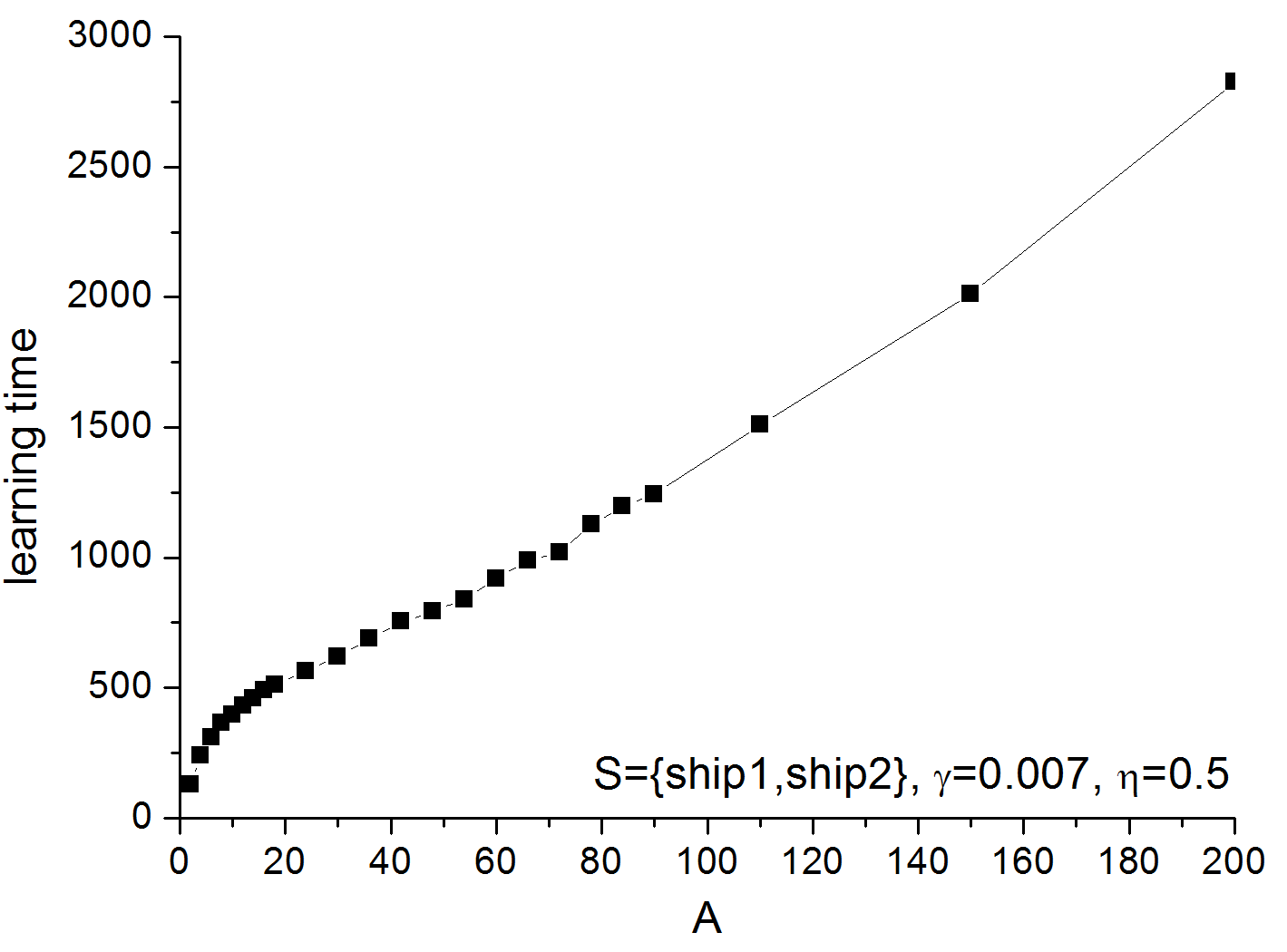}
			\end{minipage}
		\end{center}
	\caption{Performance of PS in the $n$-ship game with $A$ actions. Top: average reward as a function of played games. Middle: asymptotic reward as a function of $A$ for both a trained PS agent, shown in full black squares, and an untrained agent with equiprobable choice action depicted in empty squares.
Bottom: Learning time of PS as a function of $A$.}
	\label{fig:modifiedShipGame_PS}
\end{figure}

In Fig.~\ref{fig:modifiedShipGame_PS} (top) we show the performance of the PS agent for the 2-ship game with $A=\{2,8,16\}$. Here, each game is composed of $n=2$ time steps and the sum of all rewards given for both ships is plotted as a function of the number of games. It is seen that in all cases the asymptotic value is reached within less than 1000 games, and that more games are needed as the number of available actions $A$ is increased. A dashed line marks the maximum achievable reward of 15, and it is seen that the asymptotic averaged reward decreases when increasing $A$. Fig.~\ref{fig:modifiedShipGame_PS} (middle) shows that this decrease of the asymptotic averaged reward is approximately linear for $A \leq 200$. For each value of A the asymptotic averaged reward is given by  averaging over the last 50 games (here averaged over 1000 agents).
In Fig.~\ref{fig:modifiedShipGame_PS} (bottom) we further show the learning time of PS (calculated using 5000 agents) as a function of the number of actions, $A$. Last, we remark that a single value of $\eta$ was used for all $A$. This is because in the $n$-ship game, for each $n$ there is a single optimal value of $\eta$, irrespective of $A$, since it is the value of $n$ that determines the time length of the correlation. Here, where $n=2$, the optimal value of $\eta$ turns out to be $\eta=0.5$.

\begin{figure}[t]
	\begin{center}
		\begin{minipage}{7cm}
			   \includegraphics[width=\linewidth]{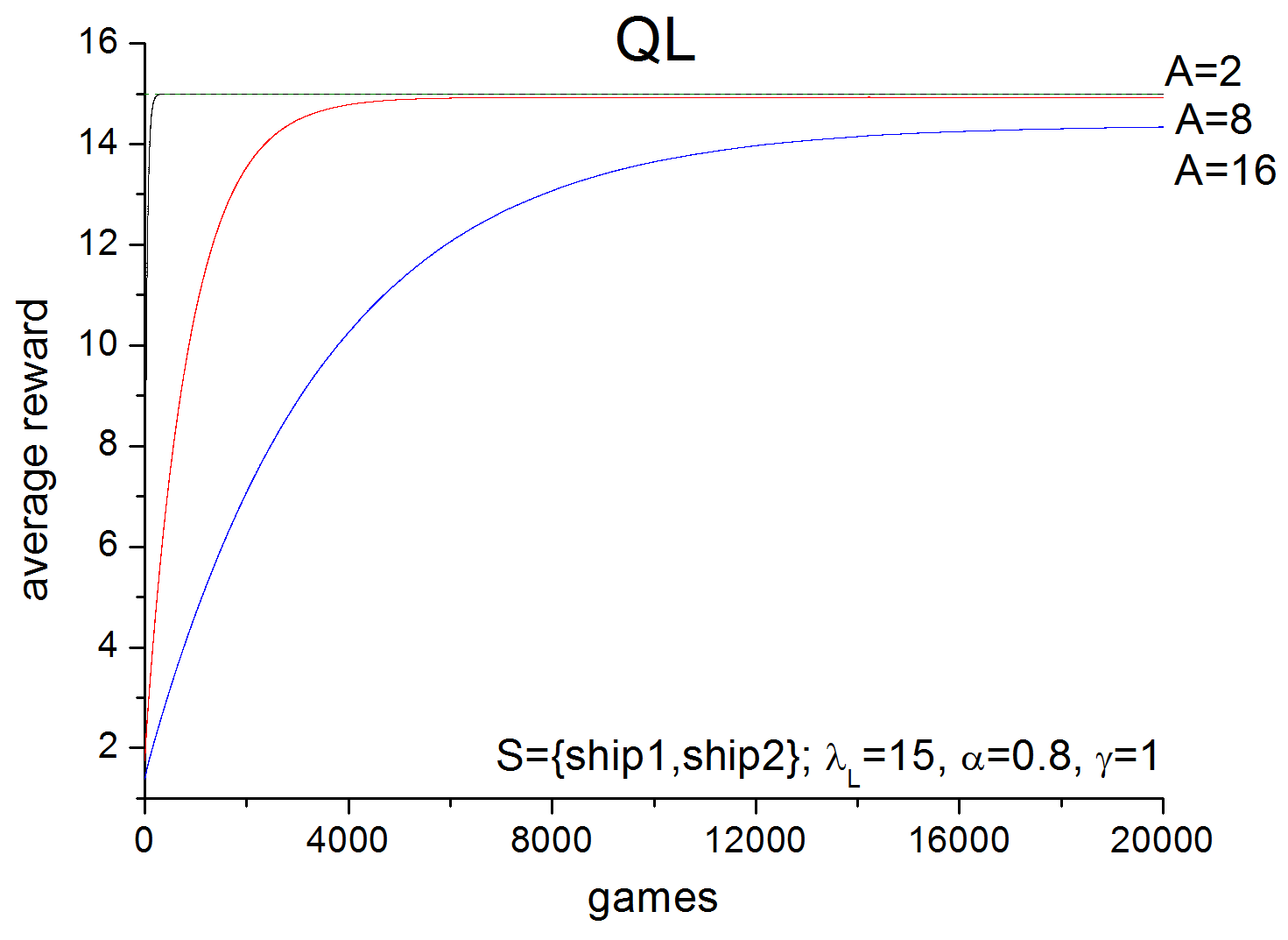}
		\end{minipage}
	\end{center}
	\begin{center}
		\begin{minipage}{7cm}
				\includegraphics[width=\linewidth]{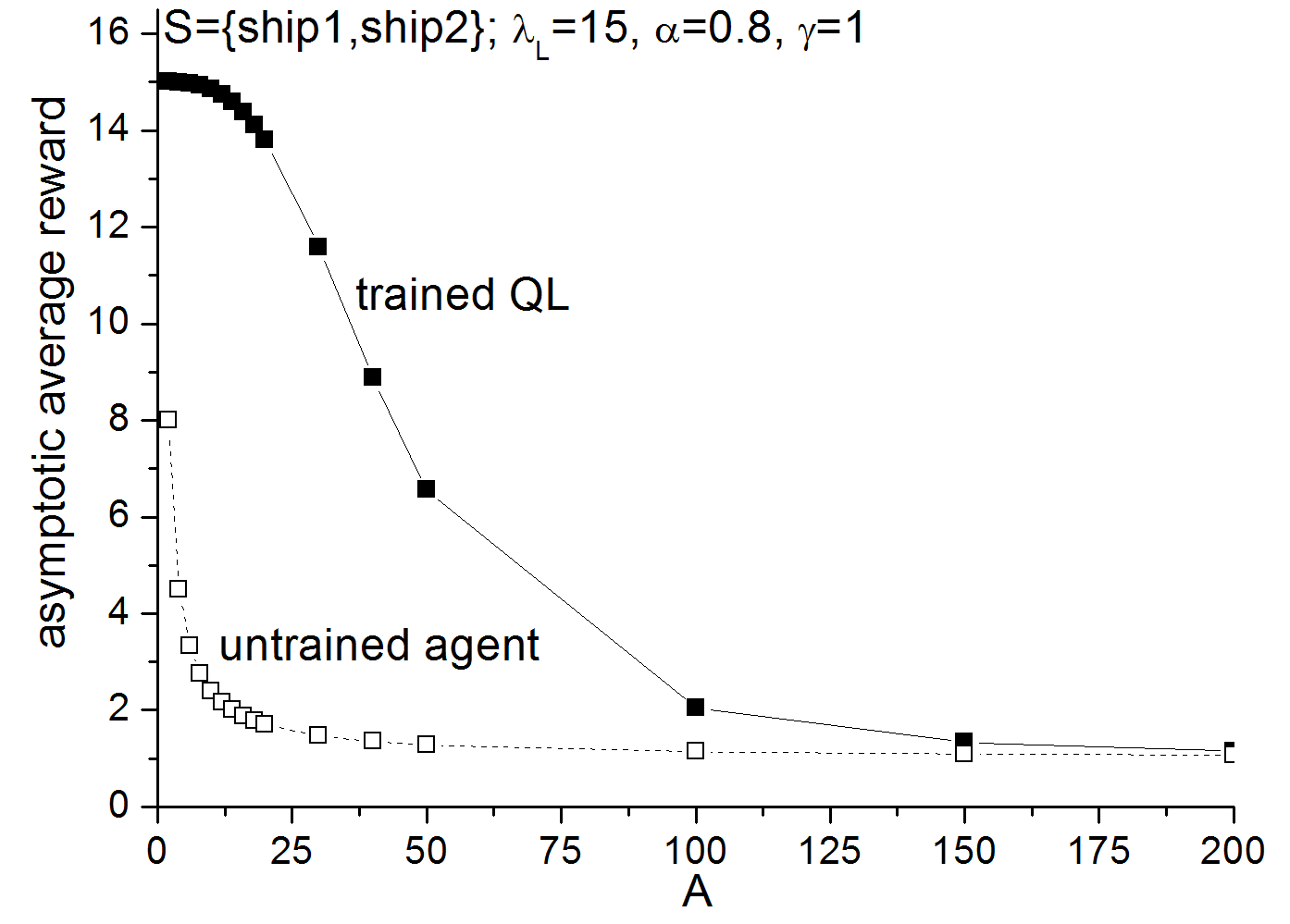}
		\end{minipage}
	\end{center}
	\begin{center}
		\begin{minipage}{7cm}
				\includegraphics[width=\linewidth]{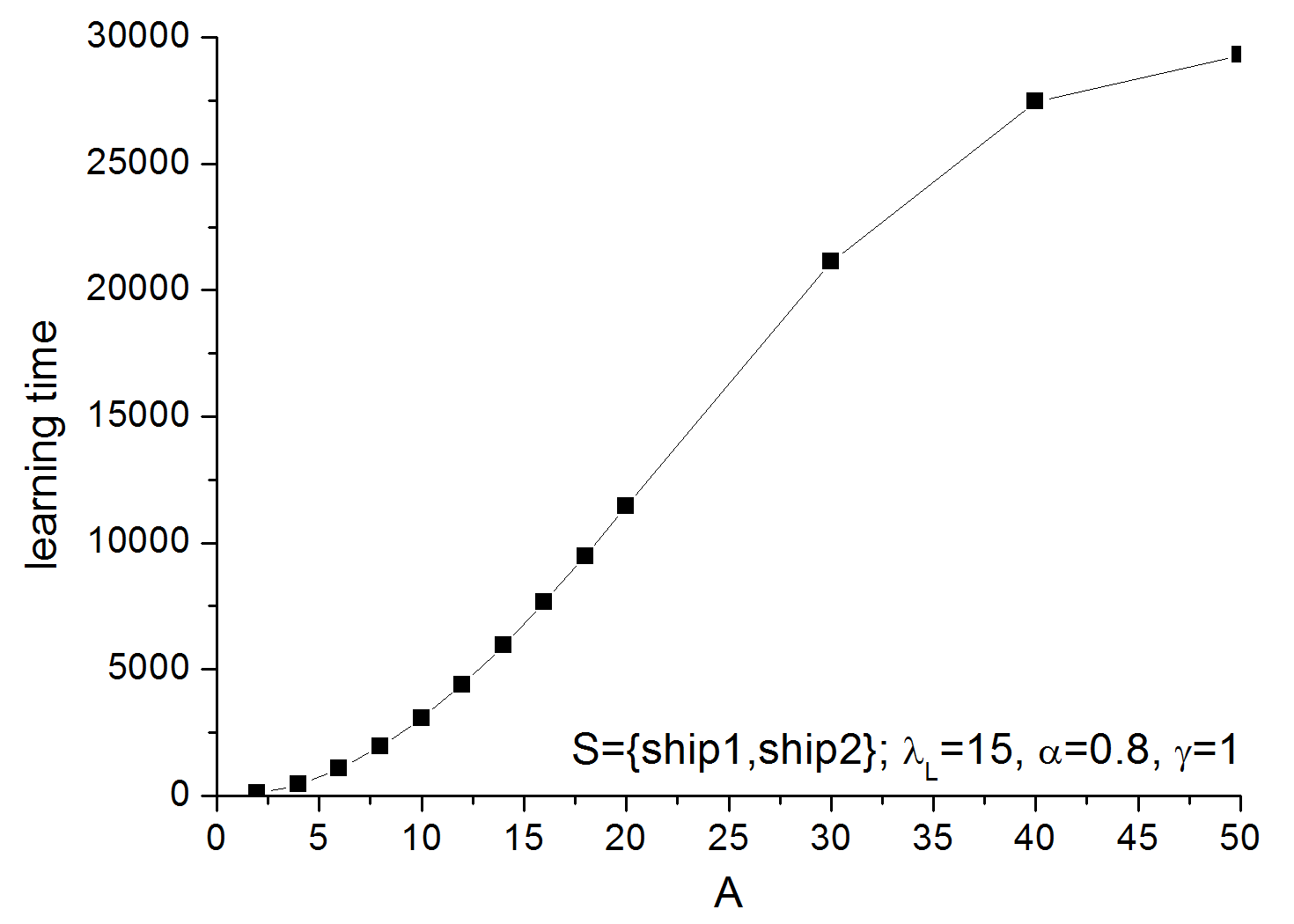}
		\end{minipage}
	\end{center}
	\caption{Performance of QL in the $n$-ship game with $A$ actions. Top: averaged reward as a function of played games. Middle: asymptotic reward as a function of $A$ for both a trained QL agent, shown in full black squares, and an untrained agent with equiprobable choice action depicted in empty squares. Bottom: Learning time of QL as a function of $A$, shown only up to $A=50$, see text.}
	\label{fig:modifiedShipGame_QL}
\end{figure}

Before ending this section we show the performance of QL for the same scenario.
Here a Q-function of 2$A$ entries was used to represent all 2$A$ state-action pairs. The values of $\alpha=0.8$ and $\gamma=1$ parameters were chosen to be (nearly) optimal ones. Fig.~\ref{fig:modifiedShipGame_QL} (top) shows the performance of QL for the $2$-ship game with $A=\{2,8,16\}$. As before, each game is composed of $n=2$ time steps and the sum of the two rewards is plotted as a function of games. The averaged asymptotic reward is shown in Fig.~\ref{fig:modifiedShipGame_QL} (middle) and the learning time is shown as a function of $A$ in Fig.~\ref{fig:modifiedShipGame_QL} (bottom).

When compared with the performance of PS as shown in Fig.~\ref{fig:modifiedShipGame_PS}, it is noted that the asymptotic reward decreases in QL rather fast. For example, at $A=50$ it is already below 7, whereas it is still above 12 for PS. As $A$ increases further the asymptotic average reward decreases toward one, i.e.\ toward a greedy behavior.
Because QL achieves very low rewards for large $A$, its corresponding learning times, as defined in section \ref{sec:scaling} are no longer indicative. For example, for $A=200$ the averaged reward of an untrained QL agent is 1.07, reached within zero time steps. For this reason
we show in Fig.~\ref{fig:modifiedShipGame_QL} (bottom) the corresponding learning times only up to $A=50$.
It is seen that the learning time of QL increases rapidly and reaches almost $3\!\cdot\! 10^4$ steps for $A=50$, whereas it is still below $10^3$ for PS. In fact, the learning time of PS does not reach $3\!\cdot\! 10^3$ steps even for $A=200$.
This implies that even though QL allows for reward to propagate backward to previous actions (see Appendix \ref{appx_QL}), it encounters difficulties when confronted with the 2-ship game with increasing number of actions.

\section{Associative memory\label{asso}}
In this section we investigate the capability of PS to exhibit a notion of \emph{associative memory}, i.e.\ to relate similar percept-clips to each other, and to use these relations to enable a more efficient learning.
This is a natural feature of PS, whose basic idea has already been introduced in the original proposal \cite{BriegelCuevas12}.
Here, we develop this idea further. 
We start with formulating the underlying mechanism of ``associative memory''. Then, we present scenarios in which efficient realization of associative memory is beneficial. We demonstrate the success of PS in such scenarios and show that it may perform even better when the mechanism of associative memory is combined with a modified excitation scheme, which we call ``clip glow''. Last, we compare the performance of PS to the one of XCS. This comparison is of value because the XCS is a method designed specifically to handle problems in which similarity in input-space may be exploited (e.g.\ in classification problems \cite{UrbanowiczMoore09}).

\subsection{Basic notion and features} \label{sec:asso-basicNotion}
\begin{figure}[htb]
	\begin{center}
		\begin{minipage}{9cm}
				\includegraphics[width=8cm]{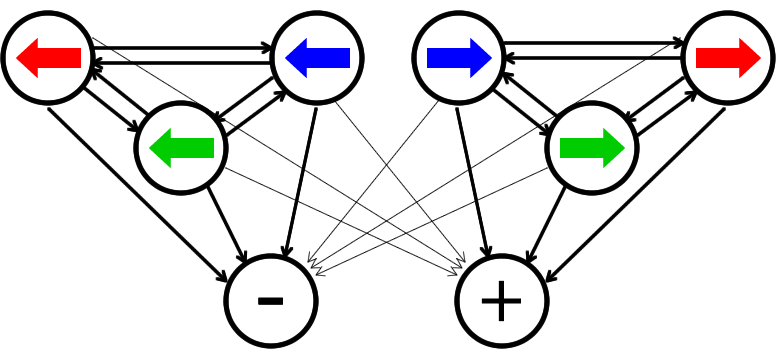}
			\end{minipage}
		\end{center}
	\caption{The clip network as built using association for the invasion game with color as additional property. Dark and light arrows indicate strong and weak edges, respectively. Associative memory is manifested via additional edges between \emph{similar} percept-clips (i.e.\ of the same shape).}
	\label{fig:associationNetwork}
\end{figure}

Within PS the notion of associative memory is realized by introducing new edges between percept-clips that are considered to be ``similar''. This is a dynamic process, where edges can be created ``on the fly'', at each time step.
A schematic visualization of a clip network after such associative memory has been built is shown in Fig.~\ref{fig:associationNetwork}. In this simple illustration, it is seen that all ``left-arrow'' (``right-arrow'') percept clips are considered similar and are therefore connected to each other (irrespective of their color).
The entire clip network is then better connected, allowing for learning to be shared between the similar percept clips.

In what follows we consider two percept clips to be similar, if they differ by exactly a single component. In addition, to avoid a situation of ``prolific association'', i.e.\ a situation in which  associative edges are built between clips whose similarity exists along irrelevant properties (such as the color property in Fig.~\ref{fig:associationNetwork}), we provide the PS with a predefined ``similarity mask", that indicates for each component in the percept space whether it is a relevant property for association or not\footnote{Using a similarity mask can be relinquished by a proper modification of the update rule, and is a subject of an ongoing work.}.

To check the performance of PS when association is enabled, we situated the PS agent in an invasion-game in which percepts are given as combinations of both shapes and colors, i.e.\ ${\cal S}\!=\!\{\Leftarrow, \Rightarrow \} \!\times\!\{\text{red, blue}\}$. At each step the agent is confronted with one of four combinations of a colored arrow and has to decide whether to go right or left. The hidden rewarding scheme is such that only shape matters, regardless of their color. Associating percepts with similar shapes but different colors might thus be beneficial.

\begin{figure}[htb]
	\begin{center}
		\begin{minipage}{9cm}
				\includegraphics[width=9cm]{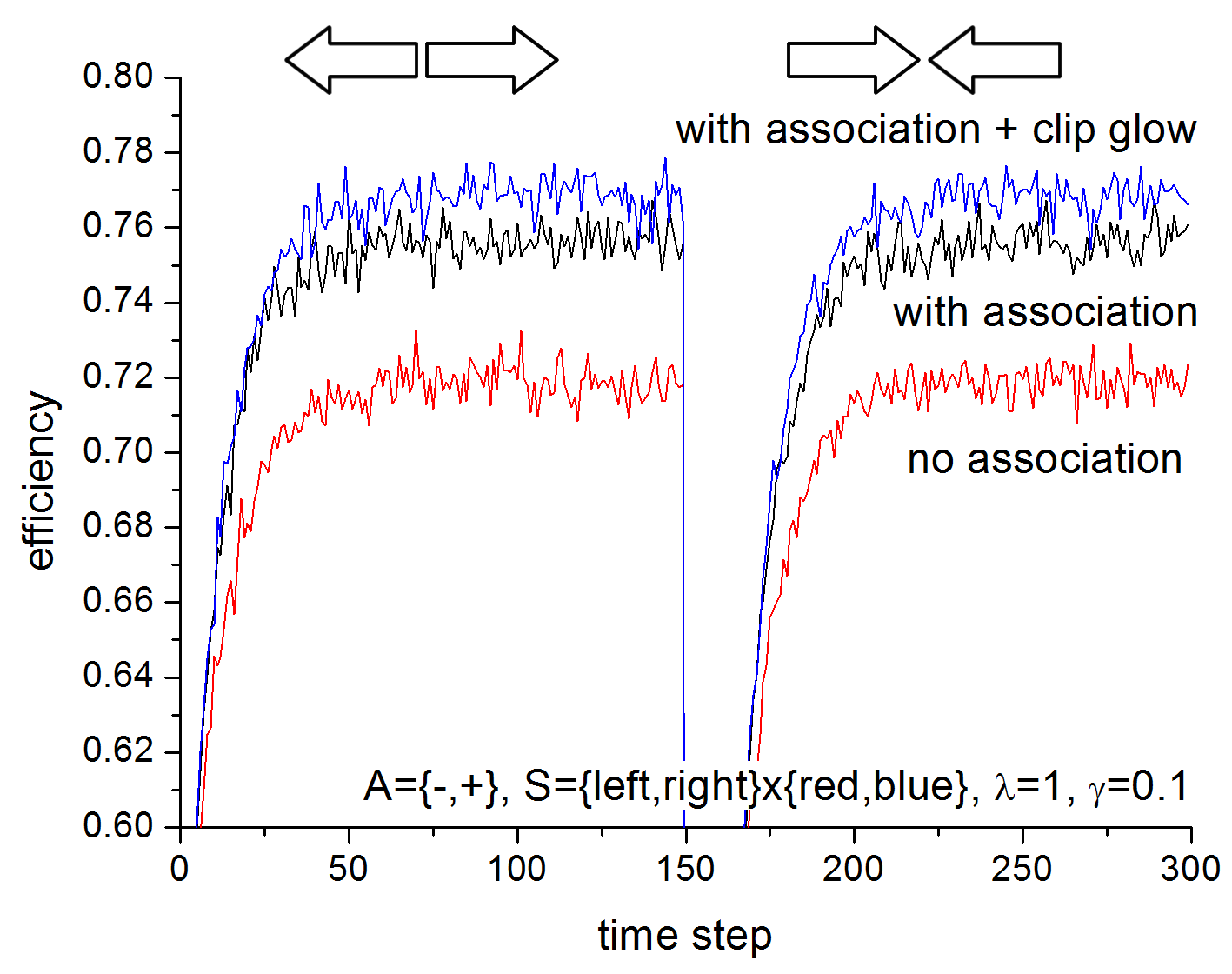}
			\end{minipage}
		\end{center}
	\caption{Efficiency of PS shown as a function of time step for the invasion game with two shapes and two colors. Here the meaning of a percept is determined only by the shape, whereas colors do not matter. 
At time step t=150, the meaning of the symbols is inverted, i.e.\ the symbol $\Rightarrow$ ($\Leftarrow$) now indicates that the attacker is going to move to the left (right). Three types of PS agents are shown: (a) without association (in red), which performs the worst; (b) with association (in black), which performs better; and (c) with association + clip glow (in blue), which performs the best.}
	\label{fig:assoColoComparison}
\end{figure}

Fig.~\ref{fig:assoColoComparison} shows the blocking efficiency of the PS agent as a function of time steps for this scenario, where at time step t=150 the meaning of the symbols is inverted, such that a right (left) arrow (colors are of no importance) then indicates that the attacker is going to move to the left (right). 
Three different kinds of PS agents are considered:
without association (in red), with association (in black), and with association + ``clip glow" (in blue). We delay our discussion on the concept of ``clip glow" to section \ref{sec:asso-clipGlow} and focus here on the performance of the first two agent types.
It is seen that, as usual (see also section \ref{sec:damping}), the asymptotic efficiency is not optimal, due to the use of a non-vanishing damping parameter $\gamma$. It is, however, interesting to note that the asymptotic efficiency is higher when association is enabled, indicating that it is indeed useful for the agent. It is further observed that the slope of the learning curve is higher when association is used, implying that the use of association allows for a shorter leaning time, as discussed below. This improvement in the performance is due to the fact that when association is enabled the length of the random walk is extended, so that \emph{more} edges may be strengthened at each time step, thereby compensating better the effect of the damping (when association is not used only a \emph{single} percept-action edge may be strengthened at a time). 
Last, we note that the better performance obtained when association is enabled, persists also when the meaning of the symbols is inverted, and has to be relearned.

\begin{figure}[htb]
	\begin{center}
		\begin{minipage}{9cm}
				\includegraphics[width=8cm]{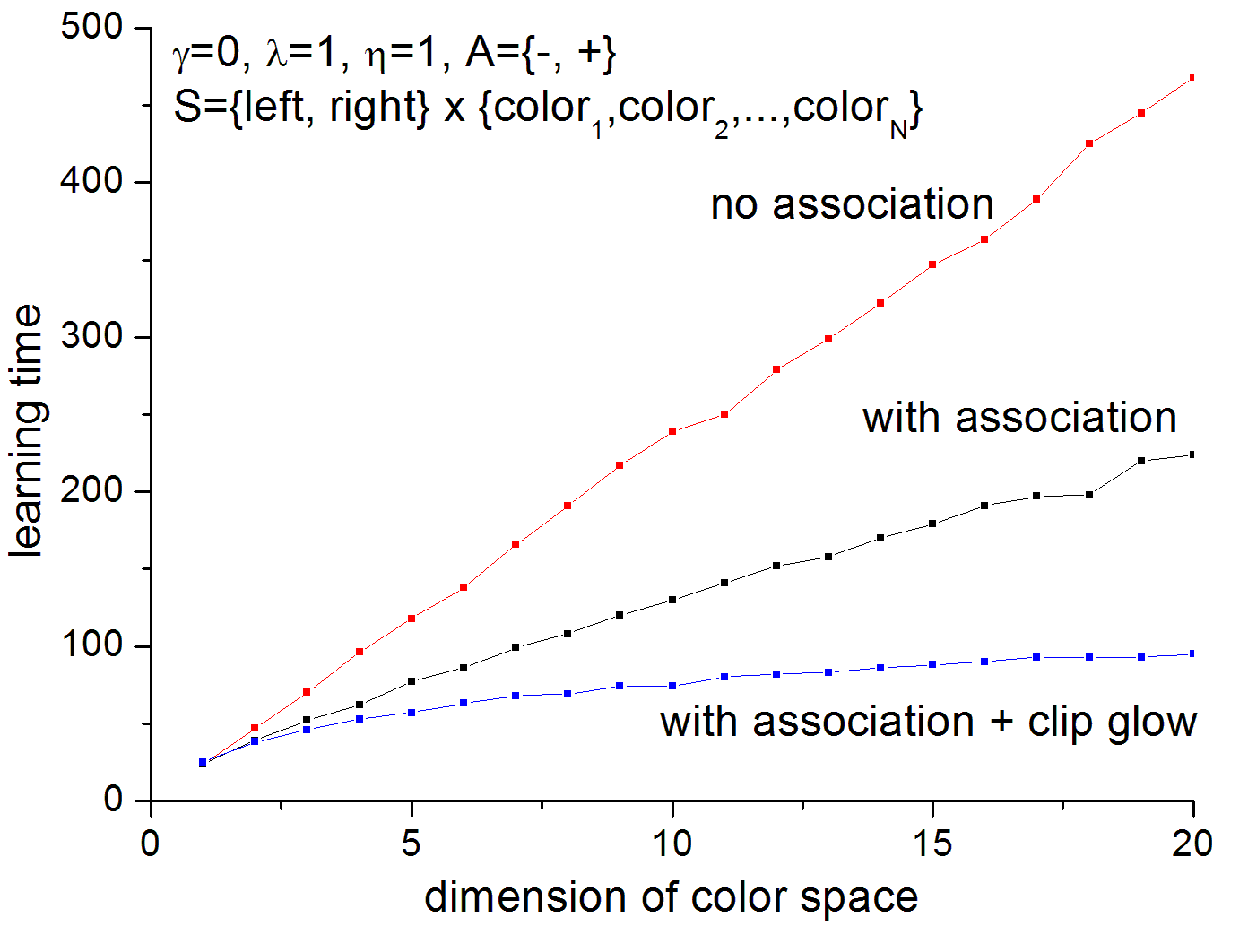}
			\end{minipage}
		\end{center}
	\caption {Learning times of PS shown as a function of the number of colors for the invasion game with two shapes and many colors. Here the meaning of a percept is determined only by the shape, whereas colors do not matter. Three types of PS agents are shown: (a) without association (in red), which learns the slowest; (b) with association (in black), which learns faster; and (c) with association + clip glow (in blue), which learns the fastest, with a scaling that is far better than linear.}
	\label{fig:assoLearningTime}
\end{figure}

It turns out that the effect of association is even more pronounced when more colors are used by the attacker (meaning that more percept-clips can be associated to each other). To see that we show in Fig.~\ref{fig:assoLearningTime} the resulting learning time as a function of color number, where shapes are kept fixed to left- and right- arrows.
The same three kinds of agents are shown as before, where we once again focus only on two kinds, namely with association (in black) and without (in red). 
It is seen that without association, the PS learning time scales linearly, whereas for agents with association the learning time scaling appears somewhat slower than linear, sufficient to exhibit a dramatic reduction in the resulting learning times. We remark that the no-association curve obtained here for
${\cal S} = \{\Leftarrow, \Rightarrow \} \times \{$color$_1$, color$_2,\dots,$ color$_N\}$ is the same learning curve (up to statistical variations) obtained in section \ref{sec:scaling}, when increasing the percept space ${\cal S}=\{$shape$_1$, shape$_2,...,$ shape$_S\}$ directly (see, e.g.\ the black curve of Fig.~\ref{fig:learningTimeS_PS}). The reason for this similarity is that when association cannot be exploited, all the percept have to be learned independently, regardless of their similarity.

The desirable performance of association comes with a price:
more edges exist in the ECM network, thus enlarging the random walk process, which effectively causes the agent to spend more time on its internal simulation before making an action.
We denote this internal time as ``deliberation time'' and
define it as the number of transitions in the clip network taken during the random walk process. When no association is exploited, the deliberation time stays constant when increasing the color space. However, when association is used, the deliberation time increases only linearly with the number of colors (not shown).
This is a positive finding for the PS, because it suggests that, at least in such scenarios, the deliberation time is always finite and that the agent never gets stuck in its ``internal world'', caught inside an endless loop.
It is, however, a good practice to set a maximal deliberation time beyond which
all processing is immediately stopped and a random action is chosen, to avoid an undesirably long simulation process. For all calculations reported in this paper, the maximal deliberation time was set to $15$, and was never reached.
We further remark that by rewarding percept-percept (association) edges stronger than percept-action edges, higher efficiency and shorter learning times may be achieved, at the price of longer deliberation times. This option was studied previously \cite{BriegelCuevas12} and will not be pursued here.

\subsection{Clip glow} \label{sec:asso-clipGlow}
In what follows we show that it is possible to boost the performance of the associative memory even further, by introducing a simple modification to the update rule of the ECM network (see Eq. \eqref{eq:updateRuleOrgI}-\eqref{eq:updateRuleOrgII} and \eqref{eq:updateEdgeRule}), which we denote as \emph{clip glow}.

The idea behind clip-glow is to strengthen not the \emph{excited edges} (i.e.\ the edges that have been used in a previous random walk process) as we have done so far, but rather to reward edges that connect \emph{excited clips}.
This is formally realized by assigning a glow parameter $g\geq 0$ to each clip (instead of edges). At the beginning, $g$ is initialized to zero, and whenever a clip is encountered, its glow parameter is set to $g=1$. The corresponding update rule is then given by:
\be \label{eq:updateRuleClipGlow}
\nn h^{(t+1)}(c_1, c_2) &=& h^{(t)}(c_1, c_2) \!-\! \gamma (h^{(t)}(c_1, c_2)\! - \! 1) \\ &\ & + \lambda g^{(t)}(c_1) g^{(t)}(c_2)
\ee
where $g^{(t)}(c_1)$, and $g^{(t)}(c_2)$ are the glow parameters assigned to clips $c_1$ and $c_2$, respectively. This means that a reward is applied to an edge if and only if it connects two glowing clips, i.e.\ clips whose $g$ values are positive. In principle, the glowing value $g$ may be damped slowly via the same damping rule given in Eq. \eqref{eq:glowDamping}. However, since ``temporal correlations" are not considered here, we let the clip-glows to completely decay after each time step, by setting the corresponding damping parameter $\eta$ to one.

An important consequence of using clip-glow is that an edge might be rewarded even if it was not visited during the random walk! According to Eq.\ \eqref{eq:updateRuleClipGlow} it is only sufficient that the clips connecting these edges were hit. Consider, for example, that during the invasion-game with the percept space given by  ${\cal S} = \{\Leftarrow, \Rightarrow \} \times \{\text{red, blue, green}\}$ the attacker shows a red left arrow. 
Assume further that the random walk takes the following sequence of clips:
\textsl{red left $\rightarrow$ blue left $\rightarrow$ green left $\rightarrow$ move left}. This sequence is rewarded because the correct ``left" action was finally reached. With the clip-glow scheme each of these clips is glowing and the reward is assigned to all of the following edges:
\bd
\item  red left $\rightleftarrows$ blue left;    \;\;  red left $\rightarrow$ move left;
\item  blue left $\rightleftarrows$ green left;    blue left $\rightarrow$ move left;
\item  green left $\rightleftarrows$ red left;   \; green left $\rightarrow$ move left;
\ed
which are five more edges than with the edge-glow scheme (where the only rewarded edges are: \textsl{red left $\rightarrow$ blue left, blue left $\rightarrow$ green left, green left  $\rightarrow$ move left, red left $\rightarrow$ move left}). These additional rewards have significant effects on the agent's performance, especially when the clip-network increases.

To demonstrate the effect of the clip-glow scheme, we show in Figs.~\ref{fig:assoColoComparison} and \ref{fig:assoLearningTime} the performance of an agent equipped with both associative memory and clip glow (in blue).
Fig.~\ref{fig:assoColoComparison} shows that the resulting asymptotic efficiency is higher, whereas  Fig.~\ref{fig:assoLearningTime} shows that learning time is then reduced even further. 
Moreover, the performance is significantly better, as the new scaling is much slower than linear: the increase of the learning time becomes slower as the color space becomes larger. This is because using the clip-glow scheme allows for even more edges to be updated and rewarded at each time step, so that the damping is compensated further.

\begin{figure}[tb]
	\begin{center}
		\begin{minipage}{9cm}
				\includegraphics[width=8cm]{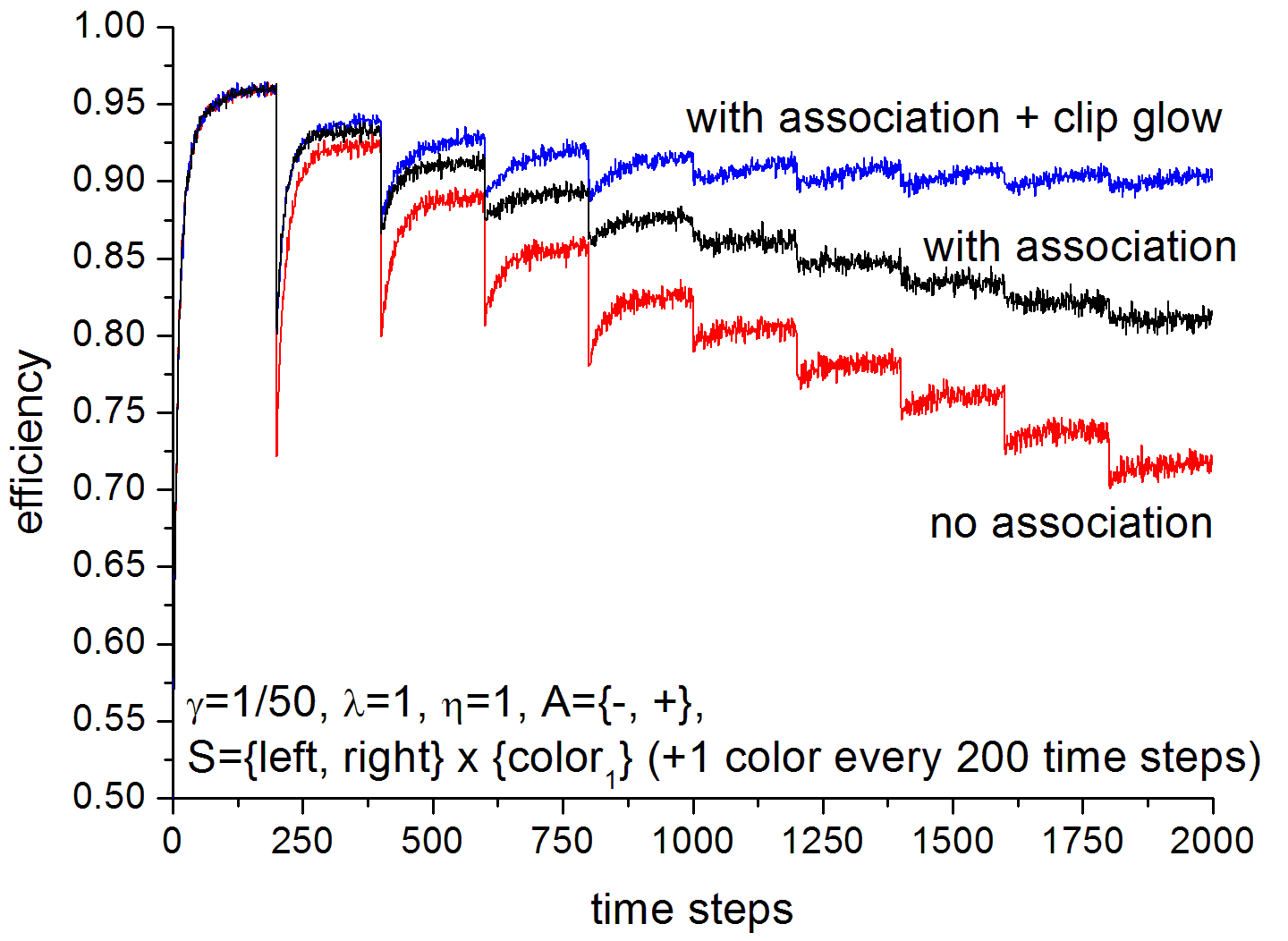}
			\end{minipage}
		\end{center}
	\caption{Efficiency of PS shown as a function of time steps for the invasion game with two shapes and many colors. Starting from a single color, a new color is added to the percept space every 200 steps. Here the meaning of a percept is determined only by the shape, whereas colors do not matter. Three types of PS agents are shown: (a) without association (in red), which performs the worst; (b) with association (in black), which performs better; and (c) with association + clip glow (in blue), which performs the best: its efficiency does not drop down (despite the increasing damping), and its required learning time for each new color becomes shorter.}
	\label{fig:assoMultiColor}
\end{figure}

To further explore the effect of combining the association scheme with the notion of clip glow, we put the agent in a scenario of ever-increasing percept space:
while the number of shapes is kept fixed (left and right arrow), the color space is increased after every 200 steps. Can the agent learn at some point that the color is of no importance? Will it learn faster due to its experience with previous different colors? Is clip glow of any use in this scenario? Fig.~\ref{fig:assoMultiColor} shows the performance in such a scenario for three different kinds of PS agents: an agent with no associative memory (in red), an agent without associative memory but without the use of clip glow (in black), and an agent with both associative memory and clip glow (in blue). The efficiency of each scheme is shown as a function of time steps. It is seen that when association is disabled, the efficiency of the agent drops with each additional color, due to the use of a non-vanishing damping parameter $\gamma$. The agent whose associative learning is enabled already does better, reaching much higher asymptotic efficiencies. Yet, when the agent can exploit the combination of both associative memory and clip glow its performance is the best.
In particular, the asymptotic efficiency hardly decreases after the fifth color and the learning time required for each new color, becomes shorter.
This desirable behavior is due to the fact that the positive effect of the associative network and the clip glow becomes more significant, the more colors there are, since learning is then shared more efficiently between similar clips.

\subsection{Comparing to XCS} \label{sec:asso-LCS} 
\begin{figure}[tb]
	\begin{center}
		\begin{minipage}{9cm}
				\includegraphics[width=8cm]{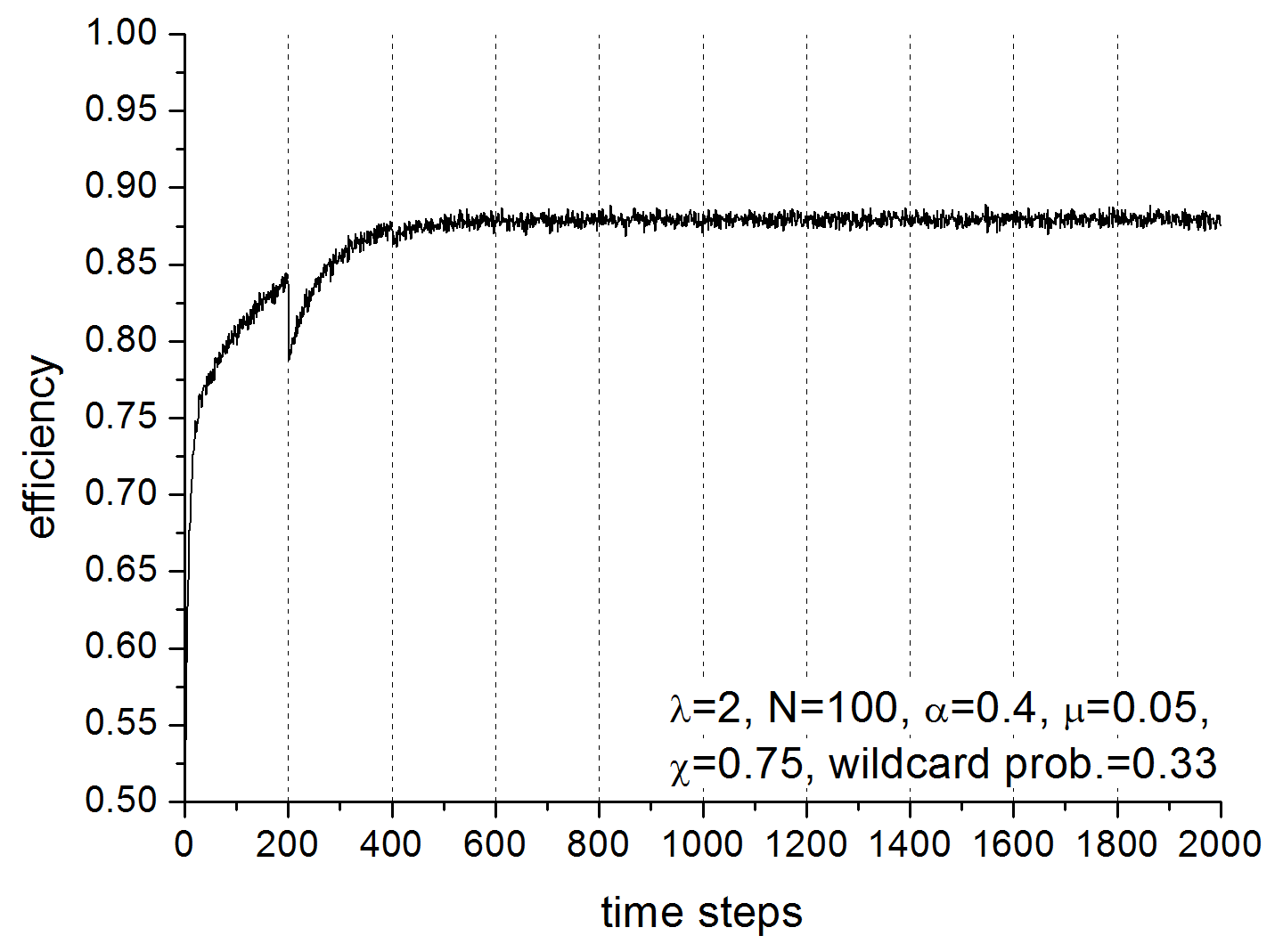}
			\end{minipage}
		\end{center}
	\caption{Efficiency of XCS shown as a function of time steps for the invasion game with two shapes and many colors. Starting from a single color, a new color is added to the percept space every 200 steps. Here the meaning of a percept is determined only by the shape, whereas colors do not matter. The reward is increased to $\lambda=2$ for achieving a reasonably high asymptotic efficiency while maintaining to use a softmax policy. For these parameters the ideal asymptotic efficiency would be $\approx 0.88$. XCS parameters (N, $\alpha, \mu, \chi$, and wildcard probability) are defined in appendix~\ref{appx_LCS}. The efficiency is the same for all numbers of colors, and the required (re-)learning time for each new color goes to zero.}
	\label{assoColorGrow_LCS}
\end{figure}

Extended classifier systems, XCS, is a particularly suitable model for scenarios of the type considered in this section, in which associating between different inputs is beneficial. This is because XCS can generalize inputs, by means of matching the inputs to conditions that use ``wildcards", as shortly described in appendix~\ref{appx_LCS}.
We thus compare the performance of PS, as shown in Fig.~\ref{fig:assoMultiColor} for the ever-increased percept space scenario, where we add an additional color every 200 time steps, to the performance of XCS under the same scenario.
Fig.~\ref{assoColorGrow_LCS} shows the resulting efficiency of XCS. It can be seen that the introduction of additional colors does not influence the asymptotic efficiency, but only temporarily disrupts the learning and partially causes a relearning for the newly added color. This leads the XCS to increasingly disregard color information in the input. By the introduction of the third color, the XCS has almost reached its asymptotic efficiency and the (re-)learning time for new colors is already negligible. This means that by this point the classifiers are general enough to handle all colors by means of wildcards. We thus conclude that in this scenario the PS performs worse than XCS for the lack of a notion of a ``wildcard clip".

\section{Composition\label{sec:composition}}

In this section we study the composition aspect of the episodic \& compositional memory (ECM). Composition is a dynamic process which accounts for structural changes in the ECM. 
In particular, it allows for spontaneous formation of new clips in the ECM network, created via combinations or variations of existing clips. These new clips may represent fictitious episodes that were never perceived before, thereby extending the variety of conceivable events and actions that exist in the ECM.
As a result, the network is less bounded by the actual past of the agent. This effectively allows the agent to generate alternative options to those it had so far encountered, thereby making it more capable and flexible.

The composition feature can be exploited in many different ways. In this paper, however, we focus on compositions of action clips only. For illustration, consider the problem of a 2-dimensional invasion game, where the agent moves on a 2D grid by means of horizontal and vertical motors. Assume that at first, the agent is provided only with four basic action clips representing ``left", ``right", ``down", and ``up" actions. The agent can thus block attackers that move either up/down or left/right. Assume further that the attackers can, nevertheless, go along diagonal directions and that the agent is partially rewarded when it performs one of the four basic actions toward the correct quarter of the attacker (e.g.\ goes left when the attacker goes left and up). Can the agent learn under these circumstances to go diagonal and block the attacker?

It is interesting to note that from a mechanical perspective there is no limitation, and the agent could move in diagonal directions, simply by activating both of its motors simultaneously. Nevertheless, the agent has no means to even conceive of the mere possibility of doing that, for the lack of a corresponding action clip. The composition feature then remedies this situation by allowing the creation of new action clips corresponding to diagonal motion.

There are various possibilities of defining clip merging and variation. Here we generalize the procedure suggested in the original proposal \cite{BriegelCuevas12}, where composition of 2-dimensional action clips was studied, and define composition mechanism for $M$-dimensional action clips, as formally described below:
\begin{itemize}
    \item Two action clips $c_a\!\!=\!\!(a_1, a_2,\dotsc, a_M)$ and $c_b\!\!=\!\!(b_1, b_2,\dotsc,b_M)$ are composed into new action clips if and only if: (a) Both corresponding actions were sufficiently rewarded for the same percept; and (b) Action clips $c_a$ and $c_b$ differ by exactly two components.
Formally, condition (a) implies that action clips $c_a$ and $c_b$ are both connected to the \emph{same} percept clip, with edges of large enough $h$-value, i.e.\ $h(c_0,c_a) \geq h_{th}$ and $h(c_0,c_b) \geq h_{th}$ for a given threshold value $h_{th}$.
	\item When the parental action clips $c_a$ and $c_b$ differ in their $i^{th}$ and $j^{th}$ components, the composition can result in two composed clips:  $c_{1}^{\text{new}}=(a_1,a_2,\dotsc,b_i,\dotsc,a_j,\dotsc,a_M)$ and $c_{2}^{\text{new}}=(a_1,a_2,\dotsc,a_i,\dotsc,b_j,\dotsc,a_M)$.
	\item A new clip is created only if it is not already present in the ECM.
	\item New action clips are to be connected to the corresponding percept clip $c_0$ with $h$-values that are given by the sum of the original $h$-values: $h(c_0,c_{1}^{\text{new}})=h(c_0,c_{2}^{\text{new}})=h(c_0,c_a) + h(c_0,c_b)$. In addition, the new action clips will be connected (at the receiving end), to all other percept clips with an initial $h$-value of 1.
\end{itemize}

In what follows we study the performance of composition in a 4D invasion game.
Here, the action space is composed of 4 motors $(a_1,a_2,a_3,a_4)$ where each motor $a_i$ can go forward (+1), backward (-1) or stay still (0), yielding a $3\times3\times3\times3$ action space of 81 possible actions. There are 8 actions, which we denote as \emph{basic actions}, for which only one of the motors is active (i.e.\ with value of $\pm1$).
At the beginning, the attacker shows any of eight possible symbols $\{1,...,8\}$, denoted ``basic symbols" as they indicate motions along one coordinate only. The defender is supplied with eight basic action-clips with which it can successfully block the attacker.
Consider now a scenario in which the attacker suddenly changes strategy and starts to show an unfamiliar symbol \ding{74}, which indicates its subsequent motion along the $(1,1,1,1)$ diagonal, that is, a forward movement in all dimensions. What will the agent do?

We assume that once the attacker has changed its strategy, it shows only the new ``diagonal symbols" \ding{74} in all subsequent rounds (this assumption is made for simplicity - removing it would make the learning process less transparent, yet with no significant qualitative insight). We further assume that the rewarding scheme (summarized in Table \ref{tab:composition_reward}) accounts also for partial success of the agent: first, all the basic four ``forward actions" (i.e.\ with one of the $a_i$ being $1$) are rewarded with $\lambda=1$; and second, all composite actions (i.e.\ non-basic actions), are rewarded proportionally to the sum of their motions, given by $q=a_1+a_2+a_3+a_4$. This ensures that actions that are partially successful are rewarded in direct proportionality with their success, corresponding to similar scenarios in real-life where combined sets of actions are rewarded more than their subsets, as is the case, for example, in the darts game, where the agent must aim its arrow both horizontally and vertically, on top of being close enough to the target plate.

\begin{figure}
	\centering
	\includegraphics[width=0.9\linewidth]{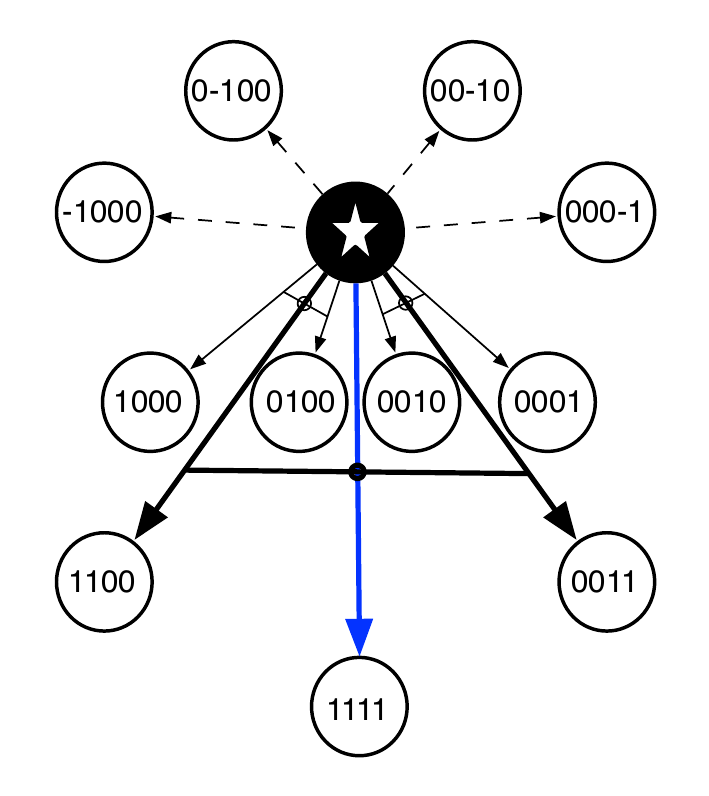}
	\caption{An illustration of part of the ECM network as built through composition in the 4D invasion game. Four kinds of arrows are shown: dashed arrows denote the weakest, unrewarded edges to the unrewarded backward basic action clips; thin black arrows denote stronger edges to the rewarded forward basic action clips; thicker black lines denote even stronger edges to partially successful composite action clips; and the blue edge is the strongest, leading to the most rewarded ``diagonal" action clip. For a composition threshold of $h_{th}>1$, only rewarded edges may be combined.}
	\label{fig:4DcompositionECM}
\end{figure}

Fig.~\ref{fig:4DcompositionECM} illustrates parts of the ECM network as built up, using composition for this rewarding scheme. Four different strengths of edges are depicted: weakest (dashed arrow), pointing to the four basic backward action clips; stronger (thin black), pointing to the four basic forward action clips; even stronger (thick black), pointing to composite partially successful action clips; and the strongest (in blue), pointing to the most rewarded ``diagonal" action clip. It is illustrated that when $h_{th}>1$, only rewarded edges are enabled to combine into new composite action clips.

\begin{figure}
	\centering
	\includegraphics[width=0.9\linewidth]{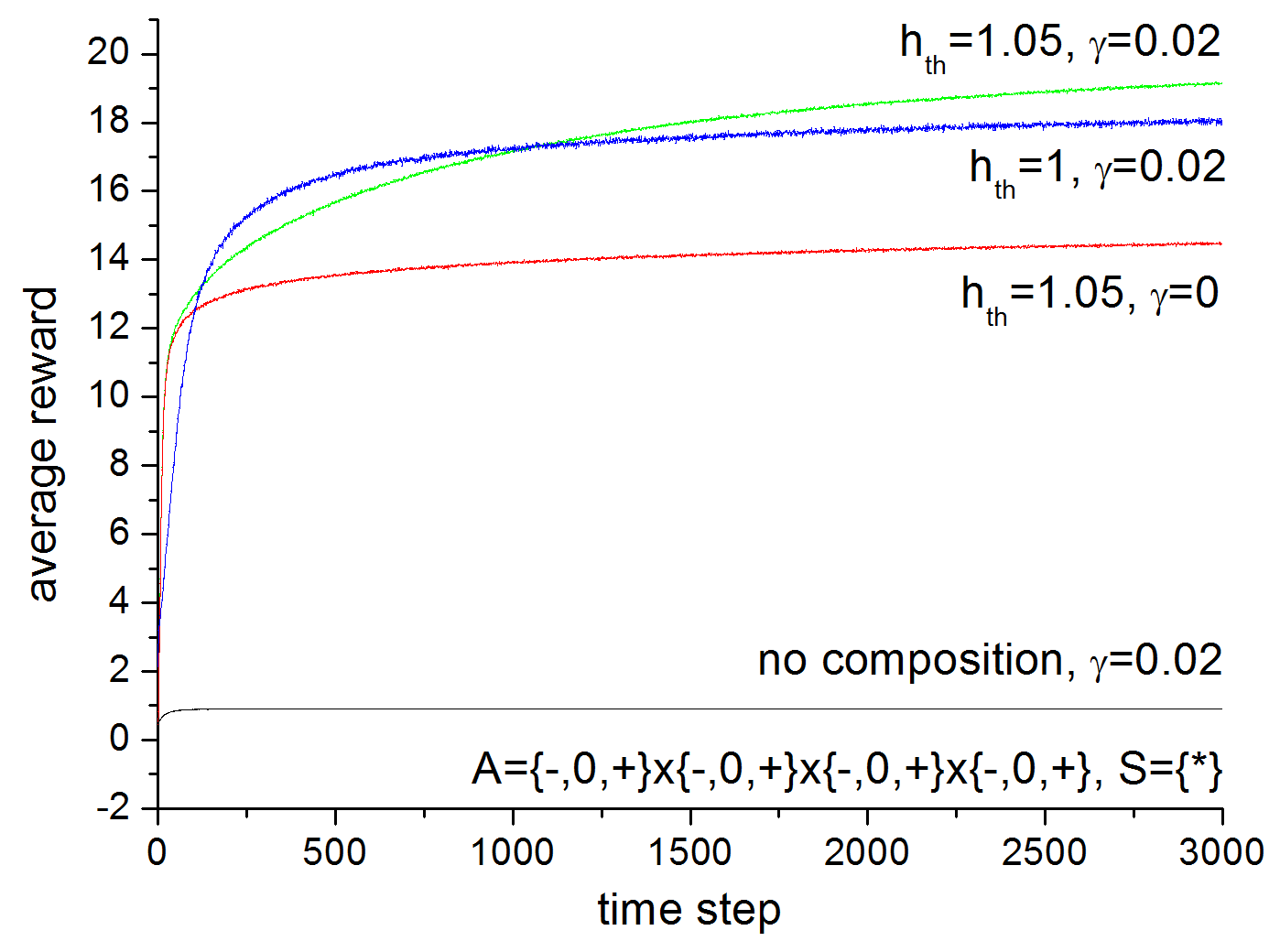}
	\caption{Average reward of PS shown as a function of time step for the 4D invasion game with only the ``diagonal'' percept \ding{74} being shown. PS agent without composition (in black) performs the worst since only basic actions with a maximum reward of $\lambda=1$ are available. Agents with composition perform better (in color). They differ by the choice of the composition threshold $h_{th}$ and the damping parameter $\gamma$:
when $\gamma=0$ (in red) the agent is effectively stuck in a local minimum, where it keeps on choosing actions that are only partially successful without ever creating the optimal ``diagonal'' action clip; using a threshold of $h_{th}=1$ (in blue) results with an immediate creation of all 81 possible action-clips, out of which 50 are never rewarded and their presence reduces the probability of finding higher rewarded actions. A higher averaged reward is depicted for $h_{th}=1.05$ (in green), where rewarded actions are combined to form potentially better composite actions.}
	\label{compositionBenefitStar}
\end{figure}

Fig.~\ref{compositionBenefitStar} shows the average reward reached by PS as a function of the number of time steps, immediately after the attacker started to show the ``diagonal" symbol \ding{74}.
When composition is disabled (in black) the agent is only able to perform the basic 8 actions, which allows for obtaining (occasionally) a maximal reward of only ~$1$, therefore resulting with a low average reward. Allowing composition with a minimal threshold of $h_{th}=1$ (in blue) results with a much higher average reward. However, using the minimal threshold of $h_{th}=1$, causes the creation of \emph{all} possible composite actions from the very beginning, that is, before the new ``diagonal" percept \ding{74} is even perceived. This excessive mode of composition has two distinct drawbacks:
first, it requires the creation of all (exponentially many) action clips, which may not be feasible for larger spaces; and second, too many unrewarded action-clips are being created, leading potentially to a non-optimal average reward.

Fig.~\ref{compositionBenefitStar} further shows that a higher average reward is achieved by setting $h_{th}=1.05$ (and $\gamma=0.02$, in green). To that end any $h_{th}=1+\epsilon$ will do. Adding $\epsilon$ prevents a default creation of ``second order" composite clips (composed of composite clips), and reduces the number of non-rewarded actions that are being composed. In fact, it turns out that when a threshold of $h_{th}=1.05$ is used, only about 10 action clips are created on average. This is a significant reduction compared to $72$ action clips that are immediately created when the minimal threshold $h_{th}=1$ is employed. In more involved scenarios, where larger action space are exploited, a selective clip-composition may become vital. We remark that a too high threshold $h_{th}$ would deteriorate the agent's performance with respect to the achievable reward (not shown), implying the existence of an optimal value for the composition threshold $h_{th}$. Fig.~\ref{compositionBenefitStar} further indicates that the desirable features of selective clip-composition when using $h_{th} > 1$ come at the cost of longer learning times. This is inevitable, as the process of selective composition is done according to the agent's experience, which takes time. This implies that there is a trade-off between higher reward, economical space usage, and
learning time. 

\begin{table}[h]
\centering
\begin{tabular}{c|c|c|c}
$q$         &            example           & $\#$equivalent clips             & $\lambda = 5q$   \\
\hline
\hline
1                           &  (1,0,1,-1)                  & $4 \choose 1$$3 \choose 1$=12    &  5               \\
2                           &  (1,0,1,0),  (1,1,1,-1)      & $4 \choose 2$+$4 \choose 1$=10   &  10              \\
3                           &  (1,1,1,0)                   & $4 \choose 1$=4                  &  15              \\
4                           &  (1,1,1,1)                   & $4 \choose 4$=1                  &  20              \\
\hline
1   (basic)                &   (0,1,0,0)                  &  $4 \choose 1$ = 4                & 1                \\
\end{tabular}
\caption{Reward scheme of the 4D invasion game, for the ``diagonal" percept \ding{74}. The reward is given by $\lambda = 5q \equiv 5(a_1+a_2+a_3+a_4)$, except for the basic forward actions which are rewarded by $\lambda = 1$. In total, there are 31 rewarded actions.}
\label{tab:composition_reward}
\end{table}

Finally, we note that when $\gamma$ is set to zero, i.e.\ when no damping is involved, the average reward drops down (in red). This is because the agent is being rewarded for actions that are only partially successful. In the absence of damping, the edges leading to this partially successful action-clips can only be strengthened but never damped. This means that the probability to choose these actions increases with time, to the extent that eventually no other actions are chosen. In particular, when the threshold is higher then $1$, the ``diagonal" action clip is never created, and the agent's performance is effectively stuck in a local minima. This is an interesting feature as it emphasizes once again the importance of damping, also for rewarded edges (see section \ref{sec:damping}).

\begin{figure}
	\centering
\includegraphics[width=0.9\linewidth]{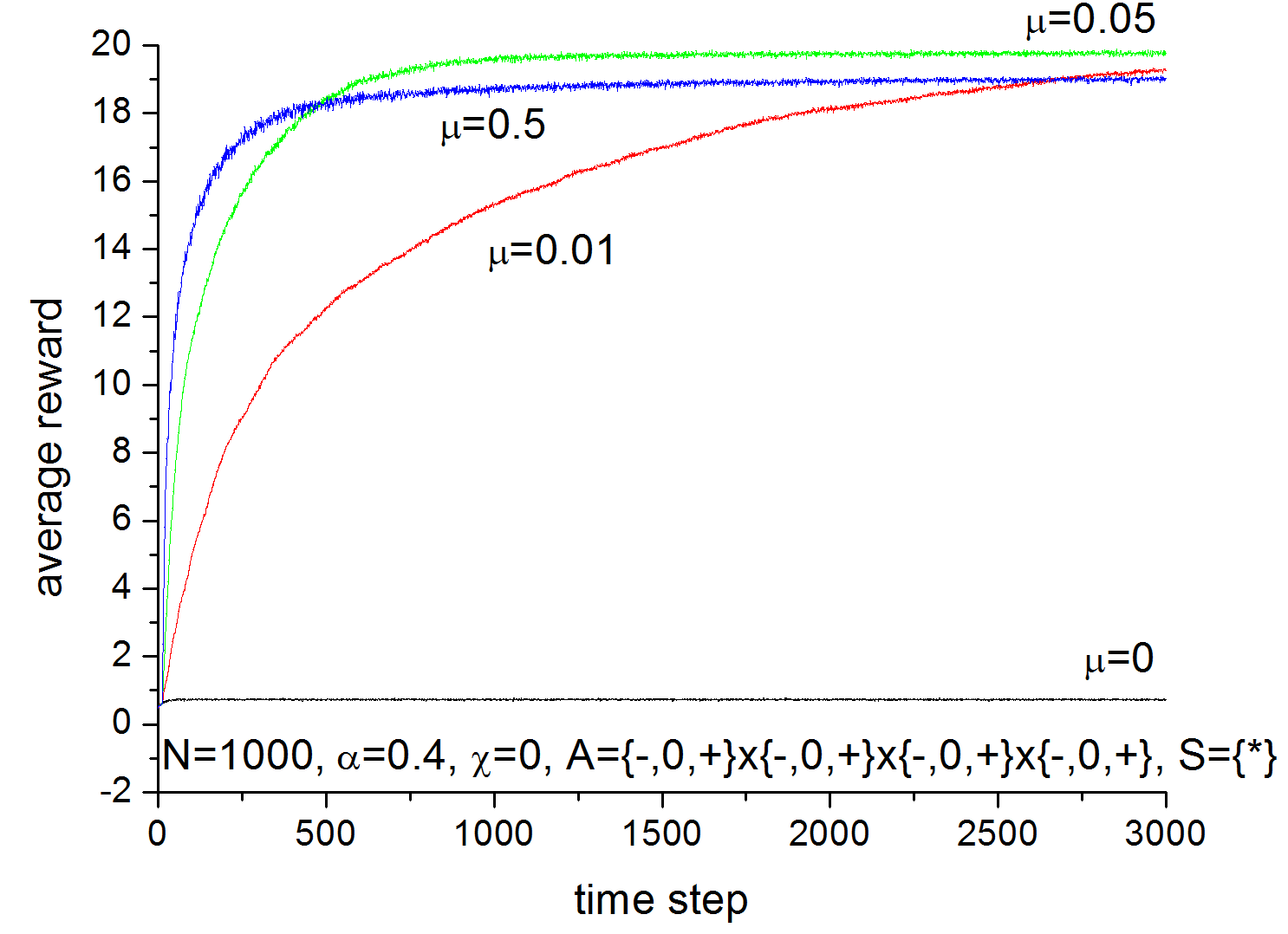}
	\caption{Average reward of XCS (the average is taken over $1000$ agents) shown as a function of time steps for the 4D invasion game with only the ``diagonal'' percept \ding{74} being shown. Without action mutation (mutation rate $\mu=0$, in black) only basic actions are available and yield an average reward below~1. Three exemplary cases for nonzero mutation rates $\mu>0$ illustrate that the creation of combined actions is possible and yields higher average rewards. A larger mutation rate causes a faster initial learning, but large mutation rates also recreate more suboptimal actions and thereby reduce the asymptotic average reward, as can be seen for $\mu=0.5$ (in blue). 
In all cases, crossover of classifiers is disabled $\chi=0$. Other XCS parameters (N and $\alpha$) are defined in appendix~\ref{appx_LCS}.}
	\label{composition2D_LCS_Star}
\end{figure}

Finally we compare the composition feature to XCS. The genetic algorithm, which is part of the XCS, is capable of performing a role similar to that of composition in the PS by mutating and combining well performing input-actions pairs, thereby creating new ones. We remark, however, that in our reference implementation, mutations are the only way allowed for creating new actions (crossover of actions is not employed because here the genetic algorithm acts on the ``action set''). Thereby, in a successful input-action pair, new composite actions can be obtained by randomly changing (mutating) the individual motor actions $a_i$ of the composite action. For the same reward scheme as used for the PS in Fig.~\ref{compositionBenefitStar}, the XCS performs similarly, with comparable learning times and efficiencies as shown in Fig.~\ref{composition2D_LCS_Star}.

Last, we remark that in the XCS the population size of classifiers is limited,
so that unsuccessful classifiers are eventually removed. This feature of removing unrewarded actions may straightforwardly be adopted to PS.

\section{Model simplicity \label{sec:comparison}}
The PS approach is distinct in its aim for a physical (and embodied) realization rather than a computational one.
Ultimately, all computational steps are to be realized by stochastic processes, and the underlying mechanism 
should be as simple as possible.

In this section we thus compare PS with the models of QL and XCS in terms of their conceptual and procedural \emph{simplicity}.
In particular, we focus on the number of parameters involved in each scheme, the basic data structures that have to be realized, and the inherent processes that should be carried out.

Starting with the number of parameters involved in each scheme, Table \ref{tab:parameters} lists all the required parameters that have to be set for PS. Here we also consider different choices that have to be made, e.g.\ ``edge-glow" vs. ``clip-glow" for PS. Although not a parameter in the strict sense, one has still to decide, which option should be used for a certain problem. It is seen that all in all PS has 6 parameters. QL is more economical  with only 3-4 parameters (see appendix \ref{appx_QL}), whereas XCS requires the tuning of about 13 parameters (see appendix \ref{appx_LCS}), out of which 6 are the most problem-dependent.

\begin{table}[h]
\centering
\begin{tabular}{l|c|c|c}
parameter                   &            range            & field              & default                \\
\hline
\hline
damping                     &  $0\leq\gamma\leq1$          & $\mathbb{R}$           &  $\gamma \ll 1$        \\
reflection time             &  $1\leq R$                   & $\mathbb{N}$           &  1                     \\
glow damping                &  $0\leq \eta\leq1$           & $\mathbb{R}$           &  1                     \\
similarity mask             &   ---                        & ---                & ---                    \\
clip- vs.\ edge-glow        &   ---                        & ---                & edge-glow              \\
composition threshold         &  $1\leq h_{th}$              & $\mathbb{R}$           &  1                     \\
\end{tabular}
\caption{List of parameters used in PS.}
\label{tab:parameters}
\end{table}

The realization of each model on a computer would require the implementation of different kinds of data structures and computational processes.
Within PS, the basic ingredient is the ECM network. In particular, a dynamic network of connected clips has to be implemented.
Two main processes are then to be realized to perform a single time step: (a) a random walk through the ECM network, to reach an action after observing a percept, and (b) updating the strength of all involved edges in the ECM, once receiving a reward. 
From a complexity point of view, the number of operations involved at each time step is of the order of the number of edges that exist in the ECM.

The QL approach relies on the Q-function, i.e.\ an array or a table, as its main data structure. At each time step
the choice of an action according to a softmax or a greedy policy has to be implemented, preferably using a binary search, and once a reward is given,  single entry in the Q-function should be updated, as well as all the state-action probabilities $P(s,a)$ for the current state $s$.
Complexity-wise these amounts to the order of $A$ operations at each time step, where $A$ is the number of available actions.

Last, the XCS scheme is based upon the availability of many (hundreds or more, for the simple problems we have studied here) individual classifiers, where to perform a single time step, all classifiers have to be checked, whether they match or not. This results in a so called  ``matching set" that is potentially large. The selection process then takes several parameters of each classifier in the matching set into account (e.g.\ fitness, predicted reward, accuracy, etc.) and chooses a single action, which determines the ``action set". When reward is assigned, the entire action set is updated, changing all the parameters of these classifiers. Afterwards, a genetic algorithm may be invoked, performing parent selection from the action set and creation of new classifiers by means of mutation and crossover. In total, each time step of XCS requires number of operations that is of the order of the number of classifiers being used. 

This simple analysis implies that while QL is a direct and economical approach, the XCS model is (by far) more involved. We argue that, from a complexity point of view, PS positions itself in between QL and XCS, yet closer to QL, in terms of simplicity and required resources. Its concept of an ECM network is more complicated and requires more space than a plain Q-function, and yet it is much simpler than the notion of classifiers.
The random walk process, that is inherent to PS, is also relatively simple and straightforward. We therefore believe that PS provides an appropriate platform for an embodied realization of an AI agent.

\section{Conclusion\label{sec:conclusion}}
In this paper we studied the model of PS, as a novel approach to artificial intelligence: we analyzed its learning features in a variety of scenarios and compared its performance with those of QL and XCS.

In our investigation we first focused on the learning features of the model, namely its asymptotic efficiency and its learning times. In particular, we were able to analytically estimate the asymptotic efficiency and the initial slope of the learning curves, for simple problems such as the invasion game. In addition, we studied the scaling behaviour of the model, where we showed that the learning time of PS scales linearly with either the input or action spaces, for the same problem.

Next we confronted the PS agent with different classes of prototypical scenarios, each of its own nature, thus demonstrating different aspects of its learning capabilities. Three types of scenarios were studied:
(a) temporal correlation scenarios, where present rewards may depend on actions done in the past; (b) scenarios for which  similarities between different percepts can be exploited; and (c) scenarios for which the environment varies, in such a way that new actions are required to maximize the reward. For each of  these three different learning classes we showed that the PS agent can reach a satisfactory level of success. In addition, for each class, we challenged the agent with learning tasks of increasing difficulty. There too, we could show that the PS agent performs well. This result is encouraging. 

It is important to note, that irrespective of specific scenario, the basic mechanism of the learning process is always the same: a random walk over network of clips connected with edges whose strength changes dynamically, due to environmental reward. This is a computationally simple, yet flexible mechanism that can be easily modified and extended. Indeed, to allow the agent to perform well in the above learning scenarios, several such extensions were made. First, we have introduced the notion of {\it afterglow}, with which edge excitations in the ECM may decay slowly, thereby allowing rewards to propagate backward in time. The PS agent can then correlate between actions of different times and achieve higher rewards in temporally correlated scenarios, that is problems of the first class. Next, we have demonstrated how associative memory, i.e.\ the ability to spontaneously create edges between similar percept clips, provides the PS agent with the ability to associate between similar percepts to speed up learning. This ability is of value for scenarios of the second class. Moreover, it was shown that combining associative memory with the notion of {\it clip glow}, according to which edges between all visited clips are rewarded, can boost the agent's performance even further. Last, the notion of {\it composition} was explored. Here, the agent is equipped with a mechanism that allows the formation of new, potentially better rewarded, action clips, by combining old ones. This formation is done dynamically, thereby providing the agent with means for surviving in varying environments, where new or more complex actions may be needed.

Throughout the paper, we compared the performance of PS with those of QL and XCS, where we put our focus on achievable asymptotic efficiencies and learning times. Initially we showed that for simple scenarios such as the invasion game, the performances of all three models are comparable. We then compared the performances of the models for each type of the above learning scenarios, where between QL and XCS, only the leading model was used as a reference.

For problems of class (a) (``temporal correlations"), all models showed a learning behavior that was  qualitatively same. Yet, important quantitative differences were noticed: First, for the $n$-ship game, both QL and XCS were relatively slow and required a significant rescaling of the reward to allow for meaningful learning. PS, on the other hand, was fast but could not achieve a complete optimal performance, due to its forgetfulness. Second, for the $n$-ship game with $A$ actions, the asymptotic reward of QL (with optimal parameters) decreased rapidly and the learning times increased very fast.
The performance of PS in this case is therefore quantitatively the best.

For problems of class (b) (``associative memory"), XCS performs better than PS. Nevertheless, PS can reach an almost optimal efficiency, despite using a non-vanishing forgetting parameter, and performs very well (better than linear) with respect to learning times. Last, problems of class (c) (``composition"), both PS and XCS are qualitatively the same. We note that for both association and composition problems, QL lacks a suitable machinery and hence cannot perform well.

We conclude that except of the case of association problems (class (b)), where XCS performs better than PS, the performance of PS is found to be comparable to the other models, and even quantitatively better as is shown for the 2-ship game with $A$ actions.
Importantly, we note that there is no single model that outperforms the other models in all cases, and that except of PS, only the XCS model is flexible enough to well-perform in all three scenarios, wheras QL is more rigid and cannot perform well in problems of classes (b) and (c).

Last, but not least, we showed that in terms of simplicity, PS is a promising model. With at most six tunable parameters, relatively primitive network structure, and a direct random walk process, it is only a bit more involved than the QL model, but much simpler than the XCS approach.

We thus conclude that projective simulation, a model for AI that aims at a physical realization in embodied agents, has its own strengths and limitations. In particular, it stands out as a competitive AI model for solving reinforcement learning problems, that is both simple and flexible at the same time.

\mbox{}
\emph{Acknowledgements:} We thank Gemma De las Cuevas for discussions. This work was supported in part by the Austrian Science Fund (FWF), through the SFB FoQuS F4012, and by the Templeton World Charity Foundation (TWCF).


\appendix
\section{Q-learning} \label{appx_QL}
Q-learning is an AI algorithm which belongs to the computational approach of reinforcement learning \cite{WAT,SuttonPhD84,Watkins89,SuttonBarto98,RusselNorvig03,Sutton90,ParrRussel97,SuttonEtAl99,Dietterich00,OrmoneitSen02,Toussaint06,SuttonEtAl08,ButzEtAl10}. 
It is an off-policy, temporal difference algorithm, in which an action-value function Q(s,a) indicates the desirability of choosing action $a$, when being in an environmental state $s$. Formally, we write $\cal{Q}: \cal{S}\times\cal{A} \rightarrow \cal{R}$, where $\cal{S}$ and $\cal{A}$ stand for all the possible states and actions, respectively.

Once the agent performs an action $a$ in a state $s$, it is given a reward $R$, and the Q-function is subsequently updated according to the following update rule:
\be \label{eq:QL_updaterule}
 Q(s^{(t)},a^{(t)}) \!\!\!&\leftarrow&\!\!\! Q(s^{(t)},a^{(t)}) +  \\
\nn &\ & \!\!\! \alpha \!\left[\!R^{(t)}\!+\!\gamma\!\max_{a^{(t+1)}} \! Q(s^{(t+1)}\!,a^{(t+1)})\!-\!Q(s^{(t)}\!,\!a^{(t)}\!)\!\right]\!,
\ee
where $t$ indicates the current time step, $0 < \alpha\leq1$ is a learning-rate parameter, which determines the significancy of new
experiences, and $\gamma$ is a ``discount factor" which determines the significancy of possible future rewards. At the begining $t=0$, all Q values are assigned a fixed value. Throughout this work, we initiallized the Q function to zero.

For each state $s$, there is a certain probability to take action $a$, that is calculated based on the Q-function. Here, we used a softmax policy \cite{SuttonBarto98}, with which the probability to take action $a$ in state $s$ is calculated according to:
\be
Pr(a_j|s)=\frac{e^{Q(s,a_j)}}{\sum_{k} e^{Q(s,a_k)}},
\ee

There exist different approaches to improve the Q-learning algorithm \cite{Lin92}.
Among them we implemented both ``Dyna-style planning" \cite{Sutton90,SuttonEtAl08} and ``experience replay" \cite{Lin92,AdamEtAl11}.
These two techniques turned out to give similar results for our basic invasion-game scenario, and therefore only results obtained with Dyna-style planning are presented in section \ref{sec:scaling}. 
In particular, our Dyna-Q implementation is based on a textbook pseudocode \cite{SuttonBarto98} (see Fig.\ 9.4 in page 233), except that we have used a softmax policy instead of a greedy one.

As implied by Eq. (\ref{eq:QL_updaterule}), QL relies on two important free parameters, namely, the learning rate, $\alpha$, and the discount factor, $\gamma$. We further regard the choice of policy as an additional parameter that has to be set. Last, using extensions such as experience-replay or Dyna introduces (at least) one more parameter (depending on the exact implementation), $P$, that is the number of replaying or planning, respectively.

\section{Learning classifier system}
\label{appx_LCS}
Learning classifier system (LCS) ~\cite{Holland75,UrbanowiczMoore09,BullKovacs05} is a machine learning model with roots in both reinforcement learning and genetic algorithms \cite{Holland75}. Within this model, percepts received from the environment are responded to by means of a set of rules, called classifiers. For a given percept all rules are examined whether they match the percept and subsequently an action is chosen from the matching rules based on the rules' payoff predictions. Individual invoked rules are modified by a reinforcement learning algorithm, and the entire set of rules is evolved by a genetic algorithm.

For the purpose of comparison with PS, we implemented a variant of the extended learning classifier system (XCS)~\cite{Wilson95}, where classifier fitness is based on the accuracy of the payoff prediction. Our implementation is based on a reference implementation by Butz and Wilson \cite{ButzWilson}, i.e.\, the original LCS proposal ~\cite{Wilson95} but with a number of improvements put forward in more recent literature. In order to facilitate better qualitative comparison with the projective simulator, some more complex features were not implemented. In particular, for our implementation the following applies: (a) The genetic algorithm is applied to the action set only; (b) For some problems (when indicated), the genetic algorithm was disabled because the classifier system contains all possible rules and the genetic algorithm would only decrease performance by destroying successful classifiers; (c) Any form of classifier subsumption algorithms were not implemented; (d) The calculation of classifier accuracy follows a power-law function; (e) The usually applied MAM-update rule (moyenne adaptive modifi\'{e}e), was not implemented in order to compare learning times on an equal footing; (f) Action selection was always done with a ``soft-policy".

For direct comparison let us briefly specify the update and action selection rules. Each classifier consists of a condition, i.e.\, a symbol or a string of symbols that is compared with the percept $s$, and an action, which should be executed if the condition matches the percept. In addition, each classifier $i$ carries a number of parameters: payoff prediction $p_i$, prediction error $\epsilon_i$, and its fitness $f_i$.
After the match set $M$ has been formed by all classifiers whose condition matches the percept, it is grouped into subsets of classifiers that propose the same action.
For each of the actions a fitness-weighted prediction is calculated from the respective classifiers by $P\!A(a_j)=\sum_i p_i f_i/\sum_i f_i$, where the sum goes over all classifiers in the match set advocating action $a_j$. Using a softmax policy for action selection, out of the existing actions in $M$ action $a_j$ is chosen with probability
\be
Pr(a_j|s)=\frac{e^{P\!A(a_j)}}{\sum_k e^{P\!A(a_k)}}.
\ee
All classifiers of the thereby chosen action form the action set $A$, which is updated and is also subject to the genetic algorithm.
For one-step problems the action set $A^{(t)}$ is updated immediately after executing the action and an immediate reward may have been received. For a received reward $r^{(t)}$, classifier parameters are updated with learning rate $\alpha$ according to
\begin{align}
p_i &\to p_i + \alpha (r^{(t)}-p_i), \\
\epsilon_i &\to \epsilon_i + \alpha (\abs{r^{(t)}-p_i}-\epsilon_i), \\
f_i &\to f_i +\alpha (\kappa_i-f_i),
\end{align}
where $\kappa_i$ is the relative accuracy of classifier $i$ with respect to all classifiers in the action set and it is calculated according to
\be
\kappa_i = \frac{ \kappa(\epsilon_i) }{ \sum_{j \in A^{(t)}} \kappa(\epsilon_j) }
\ee
with
\be
\kappa(\epsilon) = \begin{cases} \kappa_0 \left( \frac{\epsilon}{\epsilon_0} \right)^\nu &, \epsilon>\epsilon_0 \\ 1 &, \epsilon\le\epsilon_0 \end{cases}.
\ee
For multistep problems the update occurs in the action set of the previous time step $A^{(t-1)}$, i.e.\, the action set is remembered for one time step, and in addition, instead of the immediate reward the \emph{payoff} $P^{(t-1)}$ is used to update classifier predictions and errors. The payoff for $A^{(t-1)}$ is calculated from the immediate reward of that time step $r^{(t-1)}$ and the discounted prediction of the current time step:
\be
P^{(t-1)} = r^{(t-1)} + \gamma \max_{a_j \in A^{(t)}} P\!A(a_j).
\ee

The genetic algorithm is applied to the action set only if the average age of the classifiers has exceeded a given number of time steps. Upon occurrence, two classifiers are selected from $A$ with probability proportional to fitness and they are duplicated. With probability $\chi$ a single-point crossover is performed in the string of symbols that form the condition. With probability $\mu$ per symbol each of the symbols is mutated and changed. When mutating action symbols the symbol can be changed to any other symbol with equal probability. When mutating condition symbols the symbol is either changed to a wildcard, i.e.\, a ``don't care'' symbol that matches any symbol in the percept, or to the symbol given in the percept. Both classifiers are then inserted into the population. If the given maximal population size is surpassed excessive classifiers are removed from the population of classifiers with probability proportional to an estimate of the action set size that each classifier carries along and updates every time it belongs to an action set. In addition, classifiers whose fitness is below a given fraction of the average population fitness and which have been updated more than a given number of times are deleted with an enhanced probability. The exact procedure is implemented as in the reference implementation ~\cite{ButzWilson}.

In our studies the performance of the XCS depends on a number of internal parameters that need to be specified and can be tuned to optimize the performance. The most important parameters affect both learning speed, the asymptotic efficiency, and the ability of reaching a successful behavior. They are the size of the classifier population $N$, the learning rate $\alpha$, in multistep problems the discount rate $\gamma$, the number of steps after which the genetic algorithm is executed (if it is invoked at all) and the rates that govern the genetic algorithm, namely the crossover rate $\chi$ and the mutation rate $\mu$ (for the latter we specified the rate independently for conditions and actions), and the ability to generalize by specifying the probability of wildcard symbols.
In addition, there are a number of inherent parameters specified in the algorithm: (i) parameters which only affect the behavior to a small extent, such as the initial values of classifier parameters, which need to be specified but whose values are dominated in the long run by the learning rule and reward scheme, or (ii) parameters which would effectively change the algorithm like a softmax policy for action selection (rather than a greedy, $\epsilon$-greedy or roulette-wheel action selection) or (iii) parameters that are set to tested literature values and left unaltered, such as the three parameters for specifying the functional form of the accuracy function, and two parameters governing the deletion of unsuccessful but experienced classifiers.


\end{document}